\documentclass[12pt]{article}

\usepackage{amsfonts}
\usepackage{amssymb}
\usepackage{amsthm}
\usepackage[fleqn]{amsmath}
\usepackage[mathcal]{euscript}
\usepackage{mathrsfs}

\usepackage{graphicx}
\input{epsf}
\newtheorem{conjecture}{Conjecture}

\newtheorem{theorem}{Theorem}
\newcommand{\meas}{{\rm meas}}

\newcommand{\wo}{{\backslash}}

\newcommand{\Ref}[1]{(\ref{#1})}

\newcommand{\eps}{\epsilon}
\newcommand{\veps}{\varepsilon}

\DeclareMathAlphabet{\mathpzc}{OT1}{pzc}{m}{it}

\newcommand\pzcQ{{\mathpzc{Q}}}
\newcommand\pzcR{{\mathpzc{R}}}

\newcommand\pzcS{{\mathpzc{S}}}

\newcommand{\Cset}{\mathbb{C}}

\newcommand{\Nset}{\mathbb{N}}
\newcommand{\Qset}{\mathbb{Q}}
\newcommand{\Rset}{\mathbb{R}}

\newcommand{\Zset}{\mathbb{Z}}

\newcommand{\Csp}{\mathfrak{C}}

\newcommand{\Lsp}{\mathfrak{L}}

\newcommand{\Ssp}{\mathfrak{S}}

\newcommand{\cF}{{\cal F}}

\newcommand{\cM}{{\cal M}}

\newcommand{\withTtoZERO}{{\stackrel{{t\to 0}}{\longrightarrow}}}

\begin{document}

 \title{Order and Chaos in some Trigonometric Series:\\
	Curious Adventures of a Statistical Mechanic}

\vspace{-0.3cm}
\author{\normalsize \sc{Michael K.-H. Kiessling}\\[-0.1cm]
	\normalsize Department of Mathematics, Rutgers University\\[-0.1cm]
	\normalsize Piscataway NJ 08854, USA}
\vspace{-0.5cm}
\date{$\phantom{nix}$}
\maketitle
\vspace{-1.8cm} 
\begin{abstract}
\noindent
	This paper tells the story how a MAPLE-assisted quest for an interesting undergraduate problem in 
trigonometric series led some ``amateurs'' to the discovery that the one-parameter family of deterministic trigonometric 
series $\pzcS_p: t\mapsto \sum_{n\in\Nset}\sin(n^{-{p}}t)$, $p>1$, exhibits both order and apparent chaos, and 
how this has prompted some professionals to offer their expert insights.
	As to order, an elementary (undergraduate) proof is given that 
$\pzcS_p(t) = \alpha_p{\rm{sign}}(t)|t|^{1/{p}}+O(|t|^{1/{(p+1)}})\;\forall\;t\in\Rset$, 
with explicitly computed constant $\alpha_p$.
	As to chaos, the seemingly erratic fluctuations about this overall trend are discussed.
	Experts' commentaries are reproduced as to why
the fluctuations of $\pzcS_p(t) - \alpha_p{\rm{sign}}(t)|t|^{1/{p}}$ are presumably not Gaussian. 
	Inspired by a central limit type theorem of Marc Kac, a well-motivated conjecture is formulated
to the effect that the fluctuations of the $\lceil t^{1/(p+1)}\rceil$-th partial sum of $\pzcS_p(t)$, when properly 
scaled, do converge in distribution to a standard Gaussian when $t\to\infty$, though --- provided that $p$ is 
chosen so that the frequencies $\{n^{-p}\}_{n\in\Nset}$ are rationally linear independent; no conjecture has been 
forthcoming for rationally dependent $\{n^{-p}\}_{n\in\Nset}$.
	Moreover, following other experts' tip-offs, the interesting relationship of the asymptotics of $\pzcS_p(t)$ 
to properties of the Riemann $\zeta$ function is exhibited using the Mellin transform.
\smallskip


\end{abstract}
\vfill\vfill

%%%%%%%%%%%%%%%%%%%%%%%%%%%%%%%%%%%%%%%%%%%%%%%%%%%%%%%%%%%%%%%%%
\smallskip
\hrule
\smallskip\noindent
{\small 
Typeset in \LaTeX\ by the author. Based on the invited lecture with the same title delivered 
by the author on Dec.19, 2011 at the 106th Statistical Mechanics Meeting at Rutgers University
in honor of Michael Fisher, Jerry Percus, and Ben Widom.

Revised version of 08/18/2012. 

Accepted for publication in \textbf{Journal of Statistical Physics} (2012).

\noindent
\copyright 2012 The author. 
This preprint may be reproduced for noncommercial purposes.}
\newpage
%%%%%%%%%%%%%%%%%%%%%%%%%%%%%%%%%%%%%%%%%%%%%%%%%%%%%%%%%%%%%%%%%%%%%%%%%%%%%%%%%%%%%%%%%
%%%%%%%%%%%%%%%%%%%%%%%%%%%%%%%%%%%%%%%%%%%%%%%%%%%%%%%%%%%%%%%%%%%%%%%%%%%%%%%%%%%%%%%%%
%%%%%%%%%%%%%%%%%%%%%%%%%%%%%%%%%%%%%%%%%%%%%%%%%%%%%%%%%%%%%%%%%%%%%%%%%%%%%%%%%%%%%%%%%
%%%%%%%%%%%%%%%%%%%%%%%%%%%%%%%%%%%%%%%%%%%%%%%%%%%%%%%%%%%%%%%%%%%%%%%%%%%%%%%%%%%%%%%%%
\section{Introduction}
%%%%%%%%%%%%%%%%%%%%%%%%%%%%%%%%%%%%%%%%%%%%%%%%%%%%%%%%%%%%%%%%%%%%%%%%%%%%%%%%%%%%%%%%%
%%%%%%%%%%%%%%%%%%%%%%%%%%%%%%%%%%%%%%%%%%%%%%%%%%%%%%%%%%%%%%%%%%%%%%%%%%%%%%%%%%%%%%%%%
%%%%%%%%%%%%%%%%%%%%%%%%%%%%%%%%%%%%%%%%%%%%%%%%%%%%%%%%%%%%%%%%%%%%%%%%%%%%%%%%%%%%%%%%%
%%%%%%%%%%%%%%%%%%%%%%%%%%%%%%%%%%%%%%%%%%%%%%%%%%%%%%%%%%%%%%%%%%%%%%%%%%%%%%%%%%%%%%%%%
\vskip-.2truecm
\noindent
	Back in the 1990s when I was one of Jerry Percus' postdocs, I learned that Jerry's curiosity 
often let him explore unorthodox scientific ideas, just to see where they would lead to.
	In this vein, I take the invitation to celebrate the seminal contributions to statistical physics by 
three of its living legends: Ben and Jerry, and Michael, 
as a wonderful opportunity for me to follow Jerry's example and to take the three honorees
(and the reader) on a curious trip into the realm of deterministic chaos without 
pretending that I am motivated by a physics problem --- I am not! 
	Neither do I claim any mathematical sophistication!
	It is just an amusing story to tell, involving several actors, interesting mathematics, 
a few rigorous results, and some conjectures.

	The object of study is the one-parameter family of sine series 

\vskip-.65truecm
\begin{equation}
	\pzcS_p(t) 
=\label{eq:Sdef}
	\textstyle\sum_{n\in\Nset}^{}\sin(n^{-{p}}t); \qquad  \Re{p}>1,
\end{equation}

\vskip-.1truecm
\noindent
which converges absolutely for $t\in\Rset$; it's not in \cite{Zygmund}.
	Since  $t^{-1}\pzcS_p(t)\withTtoZERO \zeta({p})$,
		%$\frac{d^{2k-1}}{dt^{2k-1}}\pzcS_p(t)\withTtoZERO \zeta({(2k-1)p})$ for $k\in\Nset$, 
which is Riemann's Zeta function (art. VII in \cite{RiemannWERKE}),
the study of $p\mapsto \pzcS_p(t)$ for fixed $t$, when analytically 
extended to $p\in\Cset\wo\{1\}$, might be of interest to analytic number theorists. 
	However, I don't know whether this produces anything not already known about $\zeta$.
	Indeed, some relationships between $\pzcS_p(t)$ and $\zeta(s)$ which go beyond the obvious one just 
exhibited were pointed out to me by Norm Frankel and, independently, 
Steve Miller, in response to my SMM~106 talk.
	Prompted by their insights I added section 4.2.

	For the most part our attention will be on the $t$-dependence of $\pzcS_p(t)$ for real $p>1$. 
	Since $\pzcS_p(-t)= -\pzcS_p(t)$, it suffices to discuss $\pzcS_p(t)$ for positive $t$.

	Since the sine function with the shortest wavelength contained in $\pzcS_p(t)$ is $\sin(t)$, 
to which sine functions with ever longer wavelengths are being added, it is to be expected that
$\pzcS_p(t)$ is neither periodic nor quasi-periodic.
	Interestingly enough, the deterministic map $t\mapsto \pzcS_p(t)$ exhibits apparent chaos on small scales,
yet order on large ones.
	For example, here are two plots of $\pzcS_2(t)$: %, the first one for $0<t<500$:

\epsfxsize=9cm
\epsfysize=6cm
\centerline{\epsffile{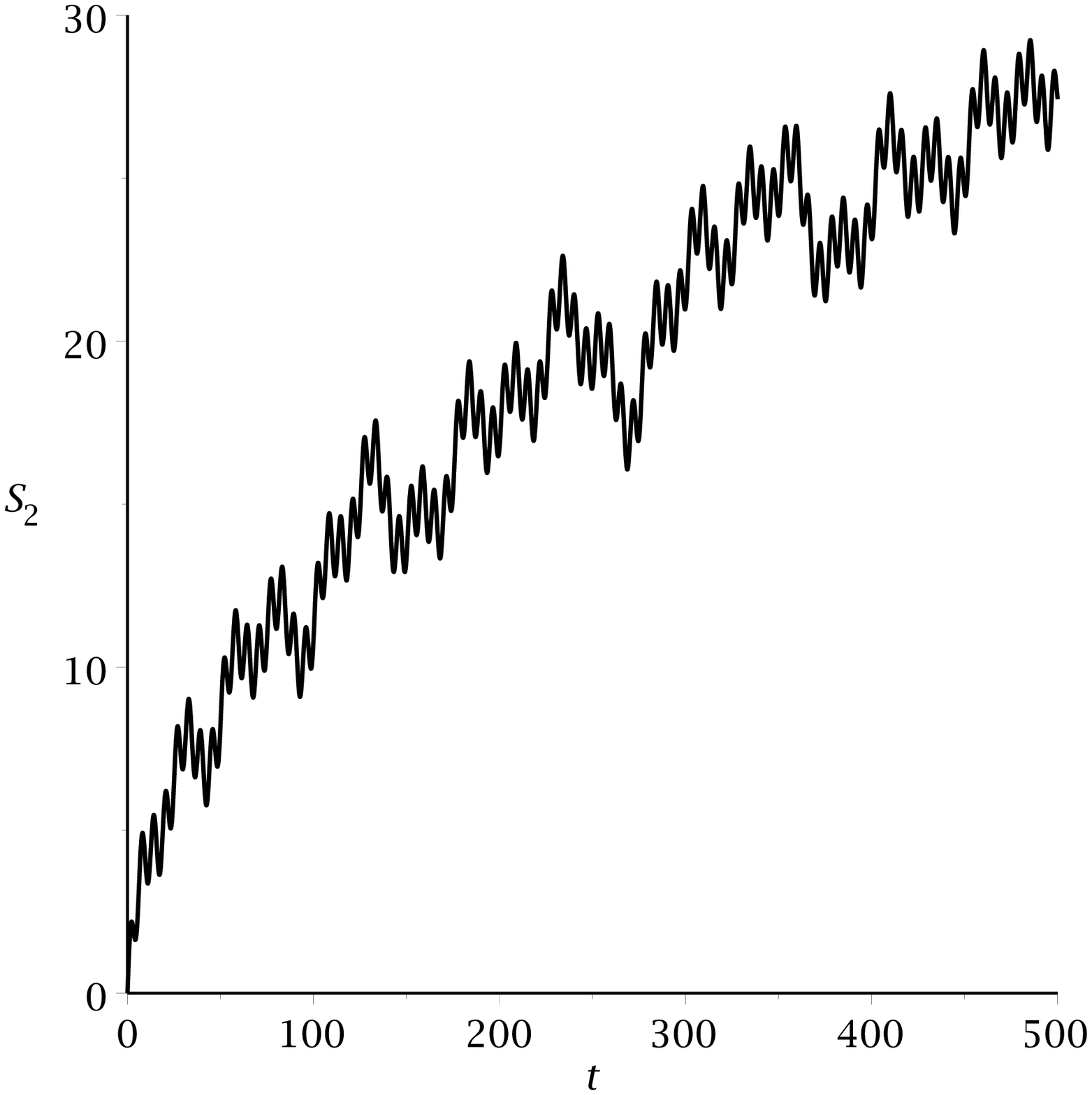}}
\begin{center}
\vskip-.5truecm
{\scriptsize{Fig.1. The 5,000-th partial sum of $\pzcS_2(t)$ versus $t$ for $0<t<500$.}}
\end{center}
\newpage

\noindent
The second one is over a 100 times larger interval of $t$ values:
\bigskip

\epsfxsize=9cm
\epsfysize=6cm
\centerline{\epsffile{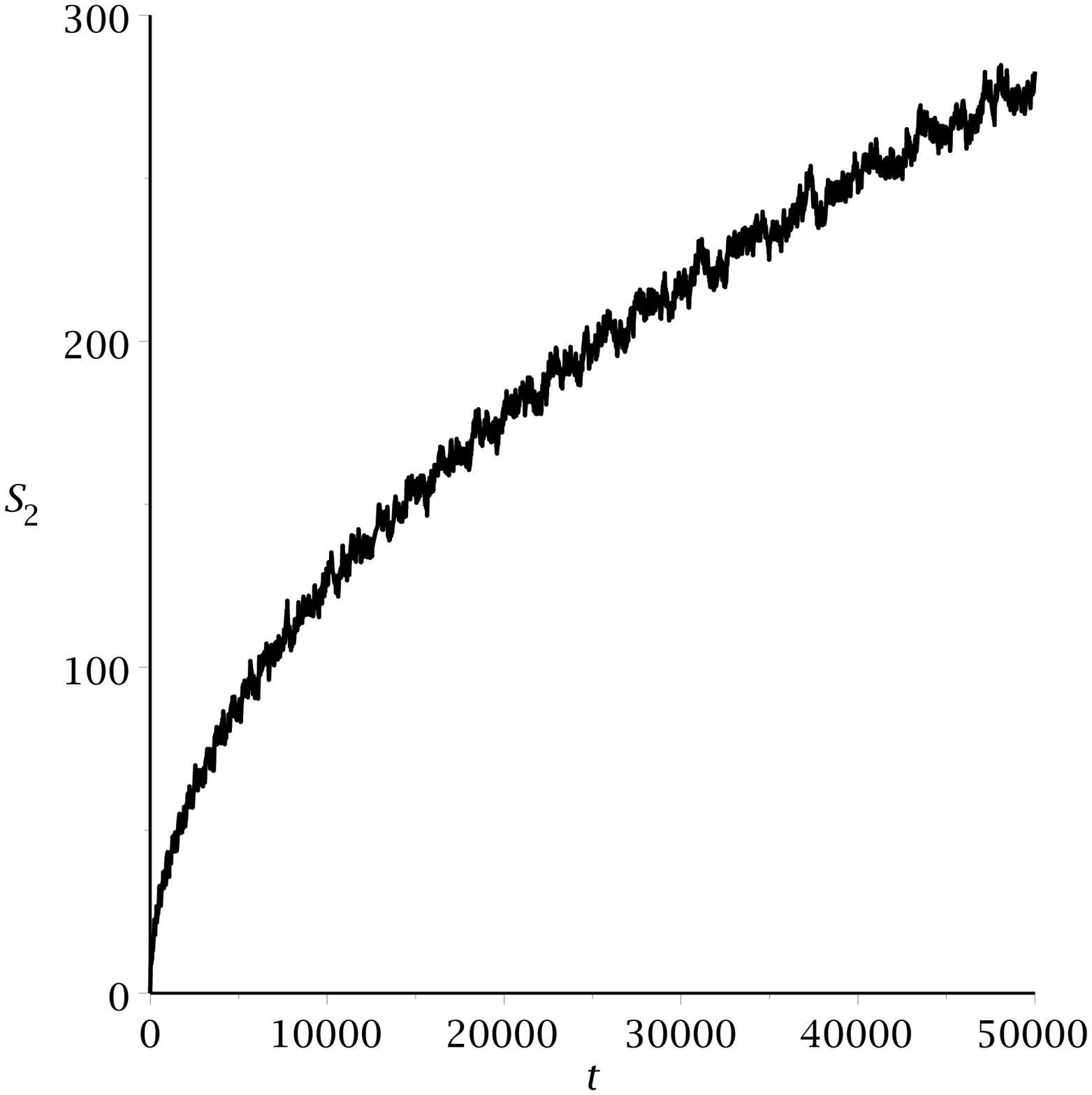}}
\begin{center}
\vskip-.3truecm
{\scriptsize{Fig.2. The 200,000-th partial sum of $\pzcS_2(t)$ versus $t$ for $0<t<50,000$.}}
\end{center}

\noindent
	We see that relative to the range of values taken by $\pzcS_2(t)$, the seemingly erratic
oscillations around their local mean appear to decrease with increasing $\pzcS_2$ value range, and the graph appears
to converge onto a rightward opening parabola, an increase of the $t$ domain by a factor 100 producing an
increase of the $\pzcS_2$ value range by a factor of 10; i.e., a square root type behavior. 
	Qualitatively similar $p$-th root type trends of $\pzcS_p(t)$ can be observed for other values of $p>1$.

	One of the rigorous results to be proved in this paper, with elementary means, is that
$\pzcS_p(t) = \alpha_p^{}t^{1/{p}} + O( t^{1/(p+1)})\; \forall\; t>0$, with
 explicitly determined $\alpha_p^{}$ for all $p>1$.
	This was obtained in partial collaboration with Jared Speck.

	For small $t$, numerical evidence is given that an $O( t^{1/(p+1)})$ bound on the deviations from the
overall trend is optimal, while it becomes lousy for large times.
	As pointed out by one of the three referees,\footnote{Any similarity with the number of honorees is
		unintended and purely coincidental.}
improved bounds on the deviations from the trend for large $t$ can be obtained if the Riemann hypothesis is assumed.
	I summarize their comments in the added section 4.1.1.

	More difficult than the determination of the trend function, but also more interesting, is the analysis of 
the deterministic, yet apparently chaotic fluctuations about the overall trend.
	A discussion of Kac's central limit theorem for sine series with rationally independent
frequencies will lead us to the conjecture that the fluctuations of the $\lceil t^{1/(p+1)}\rceil$-th partial 
sum of $\pzcS_p(t)$, when properly scaled, do converge in distribution to a standard Gaussian when $t\to\infty$
--- provided that $p$ is chosen so that the frequencies $\{n^{-p}\}_{n\in\Nset}$ are rationally linear 
independent;\footnote{It is clear that $p$ must be chosen irrational. 
		However, as noted by one of the referees,  
	$p\not\in\Qset$ is not sufficient to obtain rationally linear independent frequencies of the form
	$n^{-p}$: namely, the set  $\{n^{-p}\}_{n\in\Nset}$ will be rationally linear dependent whenever
	$p=\frac{\ln a}{\ln b}$ with integers $a>b>1$, and this formula produces rational as well 
	as irrational $p$.}
no conjecture has been forthcoming for rationally dependent $\{n^{-p}\}_{n\in\Nset}$.
	The stronger conjecture that $\pzcS_p(t)-\alpha_p^{}t^{1/{p}}$ exhibits Gaussian fluctuations, 
entertained by me at the time of SMM 106, is presumably wrong, as pointed out to me by two of the
expert referees; see section 4.1.1.
	Perhaps the discussion will prompt some interested reader to work out the definitive answer
using the professionals' tools.

	Before we now plunge into the rigorous analysis of the functions $t\mapsto\pzcS_p(t)$, I owe the
reader an answer to the burning question: How come I got to dabble in the math of these sine series?
	After all, this is not my field of expertise!
	The answer is: A question by my colleague Steve Greenfield about the graph of $\pzcS_2(t)$ for $0<t<120$
originally got me started, and the rest was curiosity about the behavior of  $\pzcS_2(t)$ for later $t$, 
and some fascination with what I found. 
	So I begin with the simpler (but not so simple) behavior of $\pzcS_p(t)$ at early times.
%%%%%%%%%%%%%%%%%%%%%%%%%%%%%%%%%%%%%%%%%%%%%%%%%%%%%%%%%%%%%%%%%%%%%%%%%%%%%%%%%%%%%%%%%
%%%%%%%%%%%%%%%%%%%%%%%%%%%%%%%%%%%%%%%%%%%%%%%%%%%%%%%%%%%%%%%%%%%%%%%%%%%%%%%%%%%%%%%%%
%%%%%%%%%%%%%%%%%%%%%%%%%%%%%%%%%%%%%%%%%%%%%%%%%%%%%%%%%%%%%%%%%%%%%%%%%%%%%%%%%%%%%%%%%
%%%%%%%%%%%%%%%%%%%%%%%%%%%%%%%%%%%%%%%%%%%%%%%%%%%%%%%%%%%%%%%%%%%%%%%%%%%%%%%%%%%%%%%%%
\section{The early time behavior of $\pzcS_p(t)$}
%%%%%%%%%%%%%%%%%%%%%%%%%%%%%%%%%%%%%%%%%%%%%%%%%%%%%%%%%%%%%%%%%%%%%%%%%%%%%%%%%%%%%%%%%
%%%%%%%%%%%%%%%%%%%%%%%%%%%%%%%%%%%%%%%%%%%%%%%%%%%%%%%%%%%%%%%%%%%%%%%%%%%%%%%%%%%%%%%%%
%%%%%%%%%%%%%%%%%%%%%%%%%%%%%%%%%%%%%%%%%%%%%%%%%%%%%%%%%%%%%%%%%%%%%%%%%%%%%%%%%%%%%%%%%
%%%%%%%%%%%%%%%%%%%%%%%%%%%%%%%%%%%%%%%%%%%%%%%%%%%%%%%%%%%%%%%%%%%%%%%%%%%%%%%%%%%%%%%%%

%%%%%%%%%%%%%%%%%%%%%%%%%%%%%%%%%%%%%%%%%%%%%%%%%%%%%%%%%%%%%%%%%%%%%%%%%%%%%%%%%%%%%%%%%
%%%%%%%%%%%%%%%%%%%%%%%%%%%%%%%%%%%%%%%%%%%%%%%%%%%%%%%%%%%%%%%%%%%%%%%%%%%%%%%%%%%%%%%%%
%%%%%%%%%%%%%%%%%%%%%%%%%%%%%%%%%%%%%%%%%%%%%%%%%%%%%%%%%%%%%%%%%%%%%%%%%%%%%%%%%%%%%%%%%
\subsection{``Can you explain the tilt?''}
%%%%%%%%%%%%%%%%%%%%%%%%%%%%%%%%%%%%%%%%%%%%%%%%%%%%%%%%%%%%%%%%%%%%%%%%%%%%%%%%%%%%%%%%%
%%%%%%%%%%%%%%%%%%%%%%%%%%%%%%%%%%%%%%%%%%%%%%%%%%%%%%%%%%%%%%%%%%%%%%%%%%%%%%%%%%%%%%%%%
%%%%%%%%%%%%%%%%%%%%%%%%%%%%%%%%%%%%%%%%%%%%%%%%%%%%%%%%%%%%%%%%%%%%%%%%%%%%%%%%%%%%%%%%%

	On April 10, 2007, Herr Dr. Prof. (emeritus) Stephen Greenfield\footnote{Steve likes to
		make fun of a German convention by calling me ``Herr Dr. Prof. Kiessling;'' 
		so I assume it's only fair when I reciprocate.}
sent me the following email:

``\emph{The attached picture is a graph of the 100th partial sum of the infinite}

\emph{series whose nth term is $\sin(x/n^2)$. You are a clever person. Why does}

\emph{the graph have the ``tilt'' that it does?}''

\smallskip

\epsfxsize=9cm
\epsfysize=6cm
\centerline{\epsffile{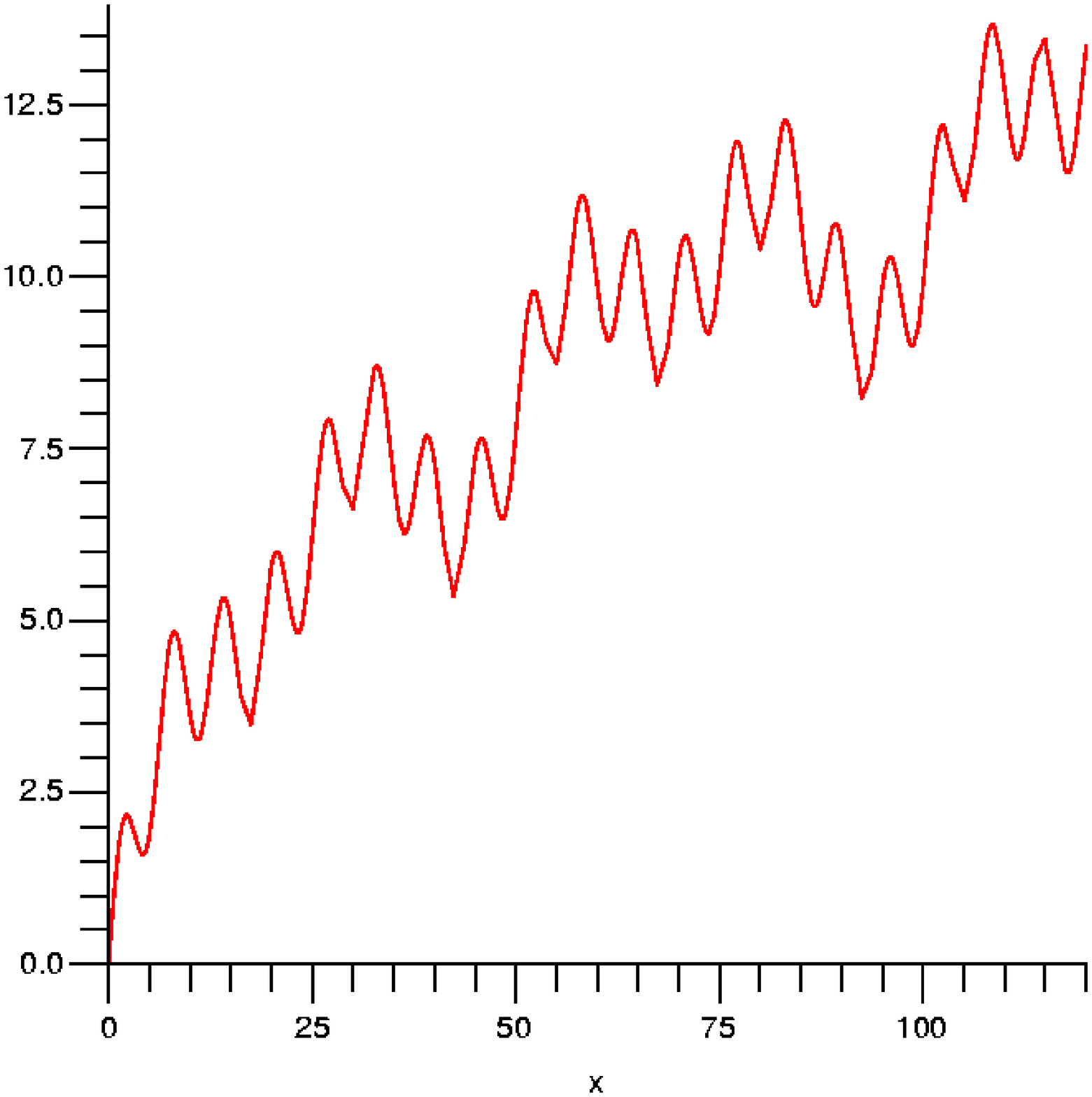}}
\begin{center}
\vskip-.3truecm
{\scriptsize{Fig.3. (Color online) Steve Greenfield's Maple plot of $\pzcS_2(x)$.}}
\end{center}

\noindent
	Nice question; but before I would drop everything I was doing at the time to rise to the challenge,
I wanted to know why he was looking at that trigonometric series.
	So I went to Steve's office three doors down the hallway to ask him what this was all about. 
	As it turned out, while trying to invent some interesting unorthodox calculus problem for his honors 
undergraduate Maple workshop, one you won't easily find solved in a solutions manual, he had played with some
unconventional trigonometric series, and this one exhibited some curious behavior: Why does the oscillating 
graph show some overall upward trend, instead of oscillating about zero, like more conventional
sine series? 
	Is there an explanation which a good undergraduate student could understand?

	An elementary, positive lower bound which tilts upward over the full domain displayed in Steve 
Greenfield's picture was soon found.
	By itself this bound does not suffice to explain the overall shape of the graph of $\pzcS_2(t)$,\footnote{I 
	am resorting to my choice of 
	variable ``$t$'' rather than Greenfield's ``$x$'' because the overall thrust of my paper is to 
	think of $t\mapsto \pzcS_p(t)$ as a deterministic process in time.}
but at least it explained why the graph of $\pzcS_2(t)$ wasn't oscillating about zero.
	More importantly, however, it would open the flood gates and let curiosity 
as to the behavior of $\pzcS_2(t)$ take hold of me, and others!
	This bound is reproduced below.

	Recall that for $t\geq 0$ we have $\sin(t/n^2)\geq t/n^2 - t^3/6n^6$. 
	For $t/n^2< 1$, this lower bound of $\sin(t/n^2)$ is off by 16$\%$ at worst.
	It can be used in the series defining $\pzcS_2(t)$ whenever
$n > \lceil{\surd{t}}\rceil$, where $\lceil{r}\rceil$ is the smallest integer not less than the real number $r$.
	Thus, writing

\vskip-.7truecm
\begin{equation}
	\pzcS_{2}(t) 
=\label{eq:SdefZWEIsplitSQRroot}
	\textstyle\sum\limits_{n=1}^{{\lceil{\surd{t}}\rceil}}\sin(n^{-{2}}t)
	+
	\textstyle\sum\limits_{n={\lceil{\surd{t}}\rceil}+1}^{\infty}\sin(n^{-{2}}t),
\end{equation}

\vskip-.2truecm
\noindent
we estimate the second sum from below by

\vskip-.7truecm
\begin{equation}
	\textstyle\sum\limits_{n={\lceil{\surd{t}}\rceil}+1}^{\infty}\sin(n^{-{2}}t)
\geq \label{eq:SdefZWEIsplitSQRrootTAYLORboundLOW}
	\Big(\textstyle\sum\limits_{n={\lceil{\surd{t}}\rceil}+1}^{\infty} n^{-{2}}\Big)t
	-
	\textstyle\frac{1}{6}\Big(\sum\limits_{n={\lceil{\surd{t}}\rceil}+1}^{\infty}n^{-{6}}\Big)t^3.
\end{equation}

\vskip-.2truecm
\noindent
	Furthermore, by the familiar Riemann sum approximations, we estimate 

\vskip-.7truecm
\begin{equation}
	{\textstyle\sum\limits_{n={\lceil{\surd{t}}\rceil}+1}^{\infty} \frac{1}{n^2}}
> \label{eq:SdefZWEIsplitSQRrootINTapproxLIN}
	\int_{\lceil{\surd{t}\rceil}+1}^\infty\textstyle\frac{1}{u^2}du 
=
	\frac{1}{\lceil{\surd{t}\rceil}+1}\,,
\end{equation}

\vskip-.3truecm
\begin{equation}
	{\textstyle\sum\limits_{n={\lceil{\surd{t}}\rceil}+1}^{\infty} \frac{1}{n^6}}
< \label{eq:SdefZWEIsplitSQRrootINTapproxCUBE}
	\int_{\lceil{\surd{t}\rceil}}^\infty\textstyle\frac{1}{u^6}du
=
	\frac{1}{5}\frac{1}{\lceil{\surd{t}\rceil}^5}\,,
\end{equation}

\vskip-.2truecm
\noindent
and so we find

\vskip-.7truecm
\begin{equation}
	\textstyle\sum\limits_{n={\lceil{\surd{t}}\rceil}+1}^{\infty}\sin(n^{-{2}}t)
\geq \label{eq:SdefZWEIsplitSQRrootAPPROXb}
	\frac{t}{\lceil{\surd{t}\rceil}+1}-\frac{1}{30}\frac{t^3}{\lceil{\surd{t}\rceil}^5}\,.
\end{equation}

\vskip-.2truecm
\noindent
	R.h.s.\Ref{eq:SdefZWEIsplitSQRrootAPPROXb} is a piecewise cubic lower bound to
l.h.s.\Ref{eq:SdefZWEIsplitSQRrootAPPROXb}, and therefore much easier to discuss than 
l.h.s.\Ref{eq:SdefZWEIsplitSQRrootAPPROXb}.
	In particular, it is easily seen to be positive and overall increasing roughly like $\sqrt{t}$;
see Fig.5 below. 
	On the other hand, the first sum at r.h.s.\Ref{eq:SdefZWEIsplitSQRroot} has just 
${\lceil{\surd{t}\rceil}}\leq 11$ terms for the $t$ (viz. $x$) interval shown in Fig.3.
	These were few enough to show by direct inspection that it
did not have enough negative terms to overpower r.h.s.\Ref{eq:SdefZWEIsplitSQRrootAPPROXb}.
	In an undergraduate class one would simply have to allude to the fact that a 
finite sum with less than a dozen terms is manageable and not go into details, though.\footnote{A suitable
	``undergraduate bound'' on $\pzcS_2(t)$ is supplied in the appendix, however.}

	Although I didn't plot it at the time, a Maple plot of the first sum at
r.h.s.\Ref{eq:SdefZWEIsplitSQRroot} shows that it itself is non-negative and tilted upward for $0<t<120$:
\medskip

\epsfxsize=9cm
\epsfysize=6cm
\centerline{\epsffile{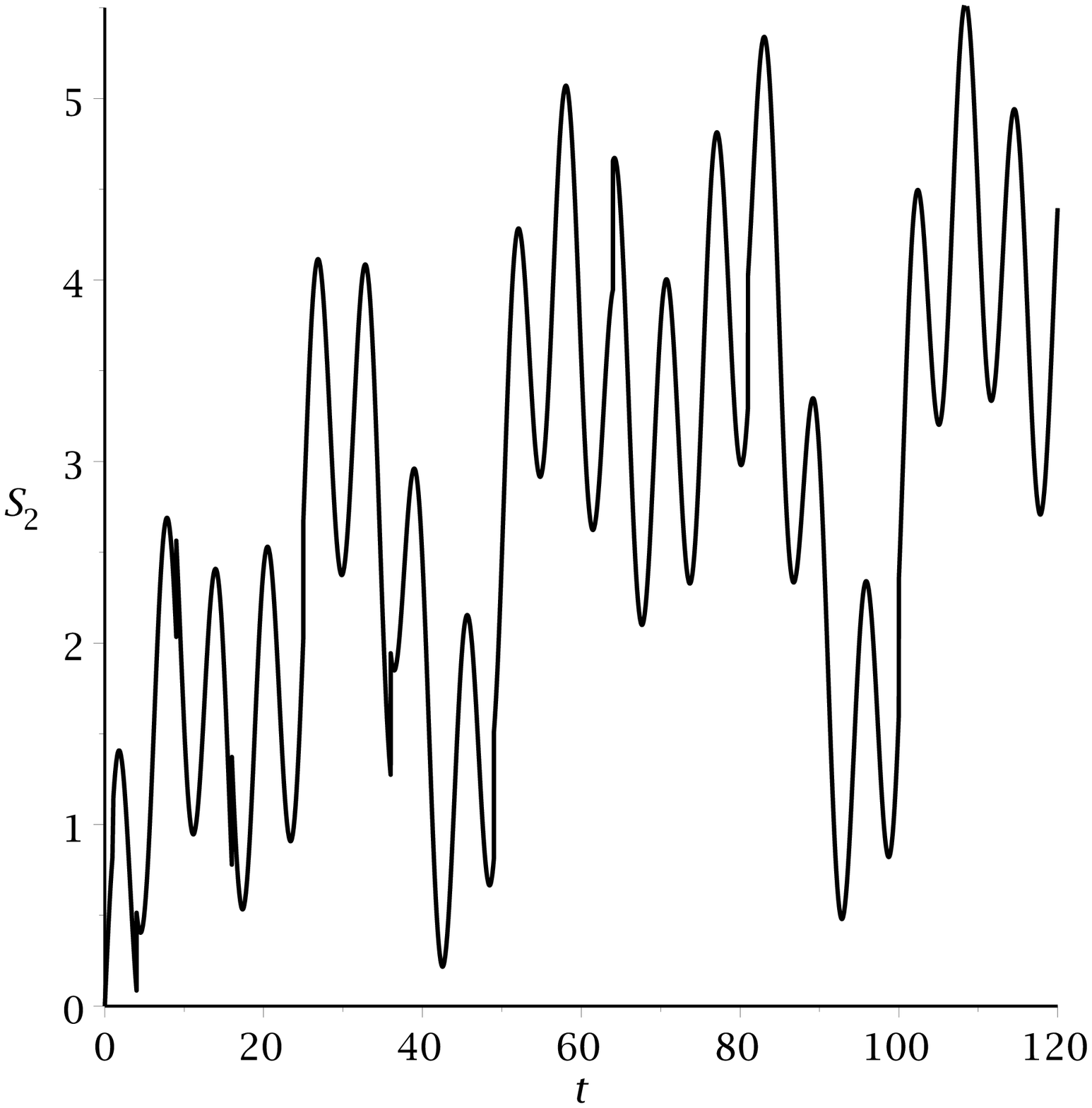}}
\begin{center}
\vskip-.5truecm
{\scriptsize{Fig.4. The first sum at r.h.s.\Ref{eq:SdefZWEIsplitSQRroot}.}}
\end{center}

\noindent
	So, curiously enough, Steve Greenfield's question applies verbatim to Fig.4!
	Yet, rather than trying to explain the overall upward tilt in Fig.4, one may want
to try to prove merely that the first sum at r.h.s.\Ref{eq:SdefZWEIsplitSQRroot} is non-negative for $0<t<120$.
	I haven't tried it, but the upshot of any such proof is: 
\emph{r.h.s.\Ref{eq:SdefZWEIsplitSQRrootAPPROXb} is an elementary lower bound to  $\pzcS_2(t)$ for $0<t<120$}.
	This bound is actually quite decent; see Fig.5 below.

\medskip

\epsfxsize=9cm
\epsfysize=6cm
\centerline{\epsffile{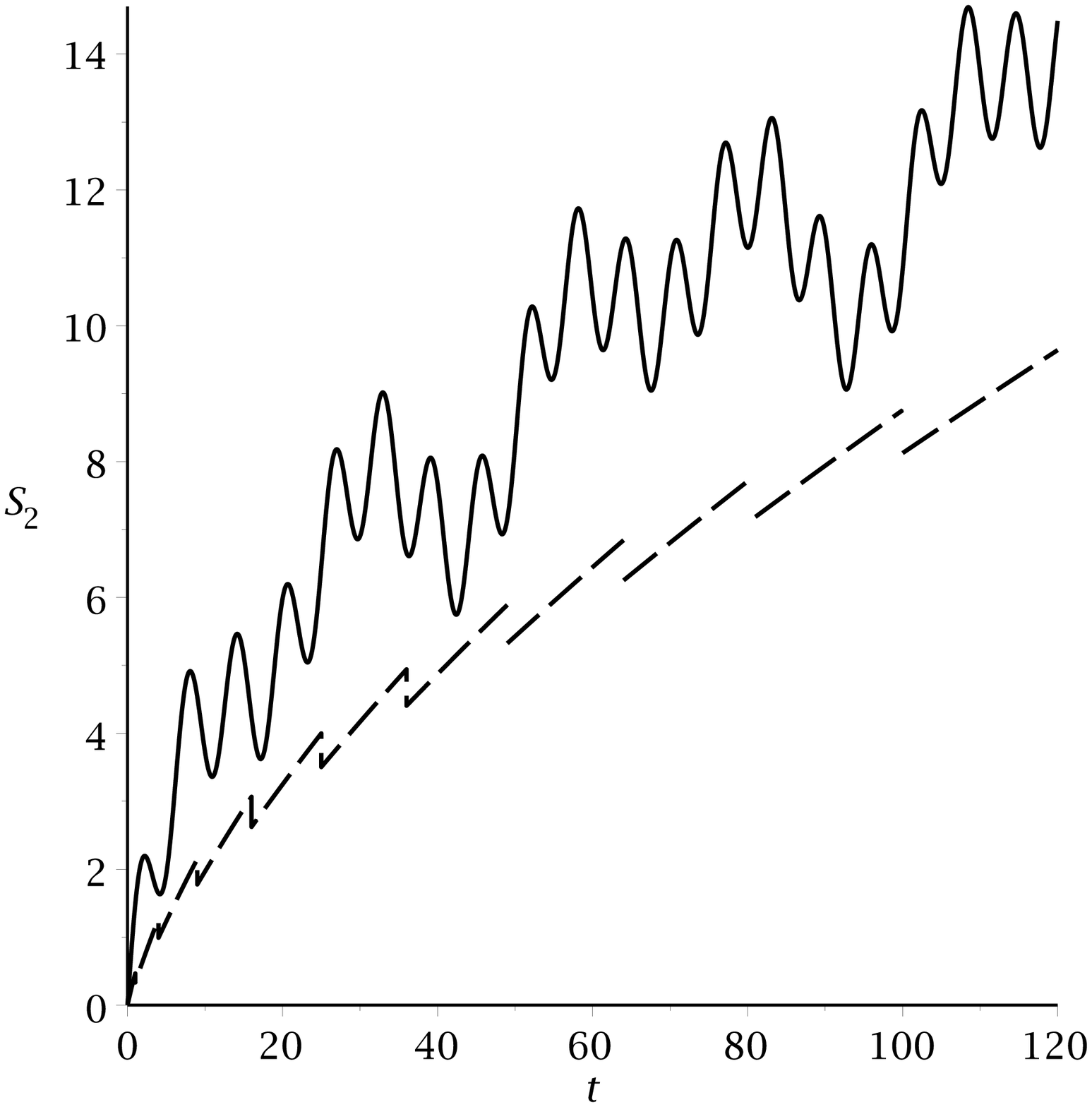}}
\begin{center}
\vskip-.5truecm
{\scriptsize{Fig.5. The 2,000th partial sum of $\pzcS_2(t)$ together with r.h.s.\Ref{eq:SdefZWEIsplitSQRrootAPPROXb}.}}
\end{center}

%%%%%%%%%%%%%%%%%%%%%%%%%%%%%%%%%%%%%%%%%%%%%%%%%%%%%%%%%%%%%%%%%%%%%%%%%%%%%%%%%%%%%%%%%
%%%%%%%%%%%%%%%%%%%%%%%%%%%%%%%%%%%%%%%%%%%%%%%%%%%%%%%%%%%%%%%%%%%%%%%%%%%%%%%%%%%%%%%%%
%%%%%%%%%%%%%%%%%%%%%%%%%%%%%%%%%%%%%%%%%%%%%%%%%%%%%%%%%%%%%%%%%%%%%%%%%%%%%%%%%%%%%%%%%
\subsection{Do all series  $\pzcS_p(t)$ have graphs like that of $\pzcS_2(t)$?}
%%%%%%%%%%%%%%%%%%%%%%%%%%%%%%%%%%%%%%%%%%%%%%%%%%%%%%%%%%%%%%%%%%%%%%%%%%%%%%%%%%%%%%%%%
%%%%%%%%%%%%%%%%%%%%%%%%%%%%%%%%%%%%%%%%%%%%%%%%%%%%%%%%%%%%%%%%%%%%%%%%%%%%%%%%%%%%%%%%%
%%%%%%%%%%%%%%%%%%%%%%%%%%%%%%%%%%%%%%%%%%%%%%%%%%%%%%%%%%%%%%%%%%%%%%%%%%%%%%%%%%%%%%%%%

	We pause briefly to inspect the early time behavior (up to $t=120$) of some sine series with other 
parameter values $p>1$.
	Here are a few examples.

	The first figure shows the graph of $\pzcS_p(t)$ for $p=\sqrt{2}$:

\smallskip

\epsfxsize=9cm
\epsfysize=6cm
\centerline{\epsffile{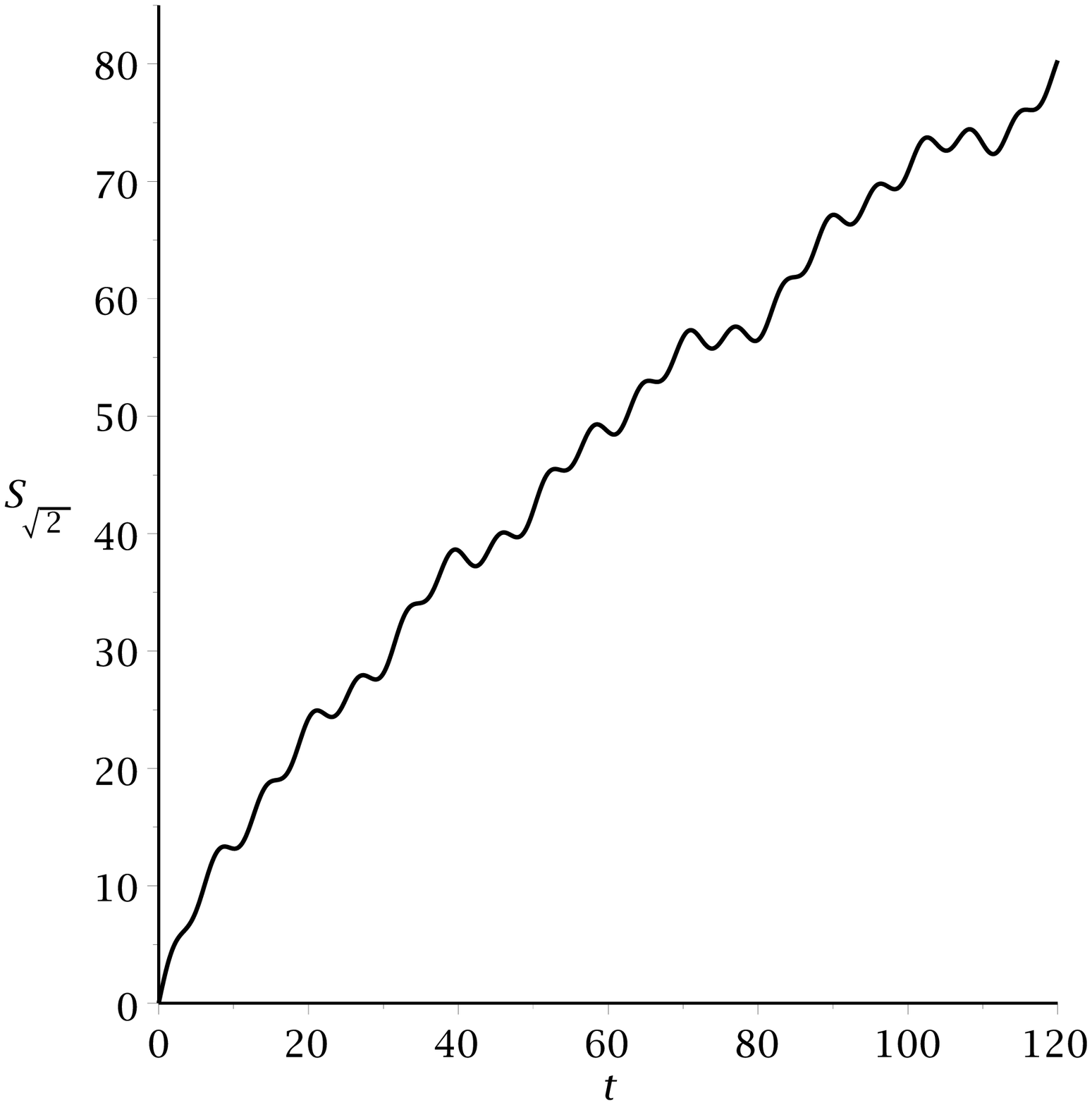}}
\begin{center}
\vskip-.5truecm
{$\qquad$\scriptsize{Fig.6. The 300,000-th partial sum of $\pzcS_{\sqrt{2}}(t)$.}}
\end{center}

\noindent
	That graph looks comparable to that of $\pzcS_2(t)$, only that the overall tilt is steeper, roughly
by a factor six.
	The amplitudes of the oscillations in the graph of $\pzcS_{\sqrt{2}}(t)$ appear smaller than in the 
graph of $\pzcS_2(t)$, but appearances are misleading, for the overall range of $\pzcS_{\sqrt{2}}$ values is 
about six times as large.
	In absolute terms the local fluctuations actually have increased, from a local amplitude of 1-2 in the graph of
$\pzcS_2(t)$ to about 2-3 in the graph of $\pzcS_{\sqrt{2}}$. 
	By the way, by ``amplitude'' I mean half the difference between a local maximum and its ensuing local minimum 
in the graph.

	Next we see the graph of $\pzcS_p(t)$ for $p=\sqrt{7}$. 
	It is plotted separately because it would show merely as a ``bottom dweller'' if incorporated in Fig.6.
\smallskip

\epsfxsize=9cm
\epsfysize=6cm
\centerline{\epsffile{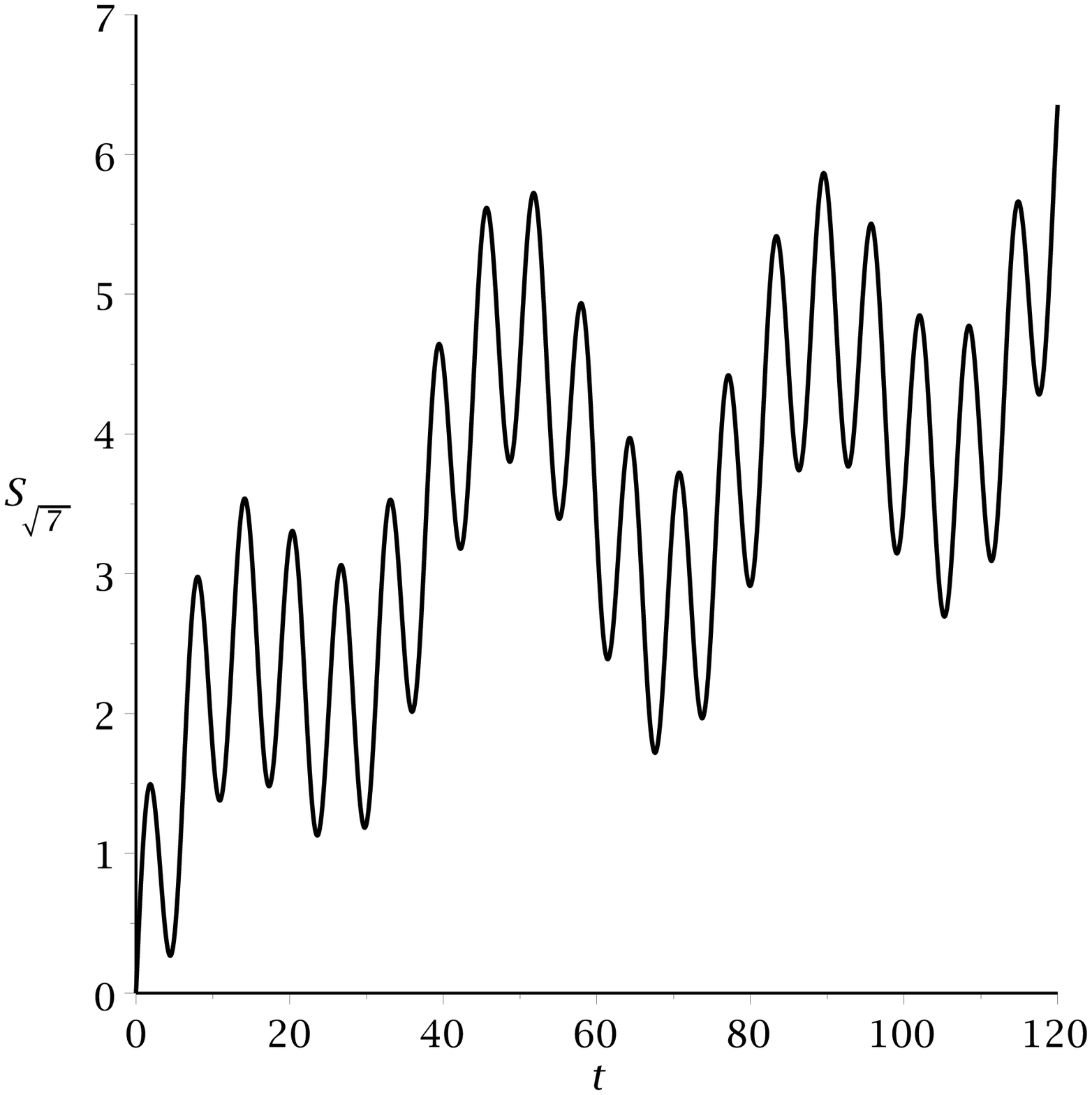}}
\begin{center}
\vskip-.5truecm
{$\qquad$\scriptsize{Fig.7. The 2,000-th partial sum of $\pzcS_{\sqrt{7}}(t)$.}}
\end{center}

\noindent
	Also this graph looks comparable to that of $\pzcS_2(t)$. 
	Now the overall tilt is less steep, roughly by a factor one half.
	The amplitudes of the oscillations in the graph of $\pzcS_{\sqrt{2}}(t)$ appear larger than in the graph of 
$\pzcS_2(t)$, but again appearances are misleading, for the overall range of $\pzcS_{\sqrt{2}}$  is about half
as large.
	In absolute terms the local fluctuations have decreased, from a local amplitude of 1-2 in the graph of
$\pzcS_2(t)$ to something closer to 1 in the graph of $\pzcS_{\sqrt{7}}$. 

	Note that the oscillations about the upward trend show $60/\pi \approx 20$ minima in all three figures, 
corresponding to the shortest wavelength involved.

	Lest the reader now thinks that, except for the magnitude of their tilts, the 
graphs of $\pzcS_p(t)$ look roughly alike for all values of $p>1$, we note that
\begin{equation}
	\lim_{p\to\infty}\pzcS_p(t) 
= \label{eq:LIMsTOinftySsOFt}
	\sin(t), \ \forall\ t\in\Rset.
\end{equation}
	Therefore, eventually the graph of $\pzcS_p(t)$ will look essentially like that of $\sin(t)$
over the whole $t$ interval $[0,120]$.
	(I spare the reader the graph of $\sin(t)$.)

	The discussion in the previous paragraph also makes it plain that the strict positivity of all displayed  
$\pzcS_p(t)$ graphs for $t>0$ is due to a too small sample of $p$ values near $p=2$.
	Eventually when $p$ is large enough, the graph of $\pzcS_p(t)$ will cross the $t$-axis.
	By comparing Fig.3 with Fig.7 it should come at no surprise that $p=3$ is large enough; however,
I didn't attempt to determine  the critical $p$-value at which the first positive solution 
to $\pzcS_p(t)=0$ occurs, nor do I know whether this would be interesting to know.

	We now turn to the more interesting problem of the overall shape of $\pzcS_p(t)$.

%%%%%%%%%%%%%%%%%%%%%%%%%%%%%%%%%%%%%%%%%%%%%%%%%%%%%%%%%%%%%%%%%%%%%%%%%%%%%%%%%%%%%%%%%
%%%%%%%%%%%%%%%%%%%%%%%%%%%%%%%%%%%%%%%%%%%%%%%%%%%%%%%%%%%%%%%%%%%%%%%%%%%%%%%%%%%%%%%%%
%%%%%%%%%%%%%%%%%%%%%%%%%%%%%%%%%%%%%%%%%%%%%%%%%%%%%%%%%%%%%%%%%%%%%%%%%%%%%%%%%%%%%%%%%
%%%%%%%%%%%%%%%%%%%%%%%%%%%%%%%%%%%%%%%%%%%%%%%%%%%%%%%%%%%%%%%%%%%%%%%%%%%%%%%%%%%%%%%%%
\section{The overall shape of $\pzcS_p(t)$}
%%%%%%%%%%%%%%%%%%%%%%%%%%%%%%%%%%%%%%%%%%%%%%%%%%%%%%%%%%%%%%%%%%%%%%%%%%%%%%%%%%%%%%%%%
%%%%%%%%%%%%%%%%%%%%%%%%%%%%%%%%%%%%%%%%%%%%%%%%%%%%%%%%%%%%%%%%%%%%%%%%%%%%%%%%%%%%%%%%%
%%%%%%%%%%%%%%%%%%%%%%%%%%%%%%%%%%%%%%%%%%%%%%%%%%%%%%%%%%%%%%%%%%%%%%%%%%%%%%%%%%%%%%%%%
%%%%%%%%%%%%%%%%%%%%%%%%%%%%%%%%%%%%%%%%%%%%%%%%%%%%%%%%%%%%%%%%%%%%%%%%%%%%%%%%%%%%%%%%%

	My tending to Greenfield's question about the ``tilt'' of $\pzcS_2(t)$ had produced the lower
estimate  to l.h.s.\Ref{eq:SdefZWEIsplitSQRrootAPPROXb} given by r.h.s.\Ref{eq:SdefZWEIsplitSQRrootAPPROXb};
this estimate is bounded below by $C\sqrt{t}$, and $\asymp (29/30)\sqrt{t}$ for $t\to\infty$.
	Moreover, in a similar fashion one obtains an upper bound 
l.h.s.\Ref{eq:SdefZWEIsplitSQRrootAPPROXb}$\leq C^\prime\sqrt{t}$ which is asymptotic to $\sqrt{t}$ for $t\to\infty$.
	These findings implied that l.h.s.\Ref{eq:SdefZWEIsplitSQRrootAPPROXb}$=C^{\prime\prime}\sqrt{t}+$small 
corrections.
	Furthermore, the first sum at r.h.s.\Ref{eq:SdefZWEIsplitSQRroot} was bounded absolutely
by $C^{\prime\prime\prime}\sqrt{t}$ and otherwise should be responsible for all the fluctuations 
visible in the plot.
	So when I presented Steve with my lower bound to $\pzcS_2(t)$, I also told him about my conjecture that 
$\pzcS_2(t)=\alpha_2^{}{\sqrt{t}}+$fluctuations for some constant $\alpha_2^{}$.

	The conjecture surprised Steve, for Fig.3 had suggested to  him that the graph of $\pzcS_2(t)$ will
continue to grow on average at roughly the same rate as the overall tilt visible in Fig.3.
	To be fair, there isn't much of an overall concave bent of the graph of $\pzcS_2(t)$ to be seen in Fig.3. 
	Using Maple, a graph of $\pzcS_2(t)$ similar to the one shown in Fig.1 was now produced, and
compared with $\surd{t}$.
	It confirmed the $\alpha_2^{}\!\surd{t}$ trend; however, $\alpha_2^{}$ had to be somewhat bigger~than~1.

%%%%%%%%%%%%%%%%%%%%%%%%%%%%%%%%%%%%%%%%%%%%%%%%%%%%%%%%%%%%%%%%%%%%%%%%%%%%%%%%%%%%%%%%%
%%%%%%%%%%%%%%%%%%%%%%%%%%%%%%%%%%%%%%%%%%%%%%%%%%%%%%%%%%%%%%%%%%%%%%%%%%%%%%%%%%%%%%%%%
%%%%%%%%%%%%%%%%%%%%%%%%%%%%%%%%%%%%%%%%%%%%%%%%%%%%%%%%%%%%%%%%%%%%%%%%%%%%%%%%%%%%%%%%%
\subsection{The pursuit of $\alpha_2^{}$}
%%%%%%%%%%%%%%%%%%%%%%%%%%%%%%%%%%%%%%%%%%%%%%%%%%%%%%%%%%%%%%%%%%%%%%%%%%%%%%%%%%%%%%%%%
%%%%%%%%%%%%%%%%%%%%%%%%%%%%%%%%%%%%%%%%%%%%%%%%%%%%%%%%%%%%%%%%%%%%%%%%%%%%%%%%%%%%%%%%%
%%%%%%%%%%%%%%%%%%%%%%%%%%%%%%%%%%%%%%%%%%%%%%%%%%%%%%%%%%%%%%%%%%%%%%%%%%%%%%%%%%%%%%%%%
	Enter Jared Speck, who at the time worked on his Ph.D. thesis research in relativity, advised
jointly by me and my colleague Shadi Tahvildar-Zadeh, and who even may have been Greenfield's TA at the time.
	When I told him about Greenfield's $\pzcS_2(t)$ and my conjecture about its $\surd{t}$-like trend, 
he didn't exactly drop whatever he was doing at the time, but the problem didn't let go of him either.
	By the end of April 11 (midnight, that is...) he had produced a conjecture as to what the constant 
$\alpha_2^{}$ could be!
	Jared noted that by boldly replacing the sum over $n\in\Nset$ with an integral over ``$dn$'' 
from 1 to $\infty$, followed by the variable substitution $\mu^2=t/n^2$, one obtains a factor $\sqrt{t}$ 
that can be pulled in front of the $d\mu$ integral, and letting $t\to\infty$ in the upper limit of that $d\mu$ 
integral, one obtains a candidate for $\alpha_2^{}$, namely
\begin{equation}
	\alpha_2^{} 
= \label{eq:alphaTWO}
	\int_0^\infty \mu^{-2}{\sin \mu^2}d\mu.
\end{equation}
	April 12 was spent pondering Jared's bold proposal.

	On the one hand, there was no reason to expect that\footnote{To avoid unnecessary confusion,
		I switch to $\nu$ rather then ``$n$'' to denote the continuous integration variable, 
		and leave $n$ to denote the discrete summation variable.}
$\int_1^\infty \sin(\nu^{-2}t)d\nu$ was an accurate pointwise approximation to $\pzcS_2(t)$ as $t\to\infty$ because
$\sin(n^{-2}t)$ hops around erratically in the interval $[-1,1]$ when $n \mapsto n+1$ for small $n$ 
(and there are more and more ``small'' $n$ as $t$ becomes large), so that one could not allude to a Riemann
sum approximation.
	On the other hand, perhaps we could show that the difference between
$\frac{1}{\sqrt{t}}\pzcS_2(t)$ and $\frac{1}{\sqrt{t}}\int_1^\infty \sin(\nu^{-2}t)d\nu$ would tend to zero,
so that his conjecture would be correct asymptotically.

	The first impulse was to resort to the splitting  \Ref{eq:SdefZWEIsplitSQRroot} 
of the series defining $\pzcS_2(t)$.
	The already collected evidence that the second term at r.h.s.\Ref{eq:SdefZWEIsplitSQRroot} 
$\asymp C^{\prime\prime}\surd{t}$, with $C^{\prime\prime}\leq 1$, suggested that all that needed to 
be done was to prove that the first sum at r.h.s.\Ref{eq:SdefZWEIsplitSQRroot} made another, though 
smaller, $\propto \surd{t}$ contribution, aside from yielding a subdominant erratic behavior.
	But there was an obstacle.
	Using the upper and lower estimates $-1\leq \sin(\xi)\leq 1$ produces upper and lower
bounds $\pm\surd{t}$ on the first sum at r.h.s.\Ref{eq:SdefZWEIsplitSQRroot} which, while compatible with
the required $\propto\surd{t}$ contribution, aren't good enough.
	There must be many near cancellations in that sum, but  a term-by-term discussion, though 
feasible for small $t$, was of course out of the question for larger $t$. 

	Later that evening I realized that the key was indeed to split the sum of $\pzcS_2(t)$ into 
two parts, but not as done in the lower estimate given in the previous section --- instead,
one had to split at some $n\propto \lceil{t^{1/3}}\rceil$ rather than at $n\propto \lceil{t^{1/2}}\rceil$.
	More precisely, with $\tau$ chosen $<\pi/2$, when $t$ is large enough then for
$n > \lceil{(2t/\tau)^{1/3}}\rceil$ any two consecutive arguments $t/n^2$ and $t/(n+1)^2$ of the sine 
functions would come to lie within a quarter period of sine; furthermore, with increasing $n$,
for fixed $t/\tau$, the consecutive arguments $t/n^2$ and $t/(n+1)^2$ would be more and more closely spaced.
	Put differently, for fixed sufficiently small $\tau$, with increasing $t$
the part of the sum of $\pzcS_2(t)$ with $n > \lceil{(2t/\tau)^{1/3}}\rceil$ will be
an increasingly better Riemann sum approximation of the integral 
$\int_{\lceil{(2t/\tau)^{1/3}}\rceil+1}^\infty \sin(\nu^{-2}t)d\nu$.
	Explicitly, if we split 
\begin{equation}
	\pzcS_{2}(t) 
=\label{eq:SdefZWEIsplitCUBEroot}
	\textstyle\sum\limits_{n=1}^{{\lceil{(2t/\tau)^{1/3}}\rceil}}\sin(n^{-{2}}t)
	+
	\textstyle\sum\limits_{n={\lceil{(2t/\tau)^{1/3}}\rceil}+1}^{\infty}\sin(n^{-{2}}t),
\end{equation}
then, for large $t/\tau$,\! by Riemann sum approximation and substitution $\mu^2\!=\!t/\nu^2$,
\begin{equation}
	{\textstyle\sum\limits_{n={\lceil{(2t/\tau)^{1/3}}\rceil}+1}^{\infty}\sin(n^{-{2}}t)}
\approx \label{eq:SdefZWEIsplitCUBErootINTapprox}
	\sqrt{t}\int_0^{t^{1/2}/\lceil{(2t/\tau)^{1/3}+1\rceil}} \mu^{-2}{\sin \mu^2}d\mu.
\end{equation}
	For any fixed $\tau$
the upper limit of the integral at the r.h.s.\Ref{eq:SdefZWEIsplitCUBErootINTapprox} grows essentially 
$\propto t^{1/6}$, i.e. it slowly but steadily diverges to $\infty$ as $t\to\infty$, and so this integral
converges to r.h.s.\Ref{eq:alphaTWO}.
	Moreover, the first sum in \Ref{eq:SdefZWEIsplitCUBEroot} is obviously subdominant.
	Better yet, this erratic term  should have plenty of near self-cancellations, and if 
one could show that it vanished on average, then even Jared's replacing of $\pzcS_2(t)$ by
$\int_1^\infty \sin(\nu^{-2}t)d\nu$ could conceivably be vindicated in an average sense.
	In any event, by now I had become convinced that Jared's conjecture for $\alpha_2^{}$ was right, 
and I sent an email to him and Steve detailing my thoughts.
%	Confident that the asymptotics was now basically a done deal, I went to bed. 

	\hfill An hour or so later, but still the same day  (almost midnight, again), Jared 
replied with the following email (temporarily we are back to $x$ instead of $t$):\footnote{For 
	convenience of the reader I use \LaTeX\ to display formulas which Jared
	described in his email.
	I deliberately left the amusing typo (which happens if you work until midnight!)}

\bigskip

``\emph{Hello guys.  Using Maple,  I summed the first 200,000 terms and graphed}

\emph{this partial sum from $x=0$ to $x= 50,000$.  I also graphed, in yellow, $C\sqrt{x}$,}

\emph{where $C = \int_0^\infty u^{-2}\sin u^2du$ (accurate to 8 digits). Of course, in principal,}

\emph{the computer could be making all sorts of round off errors, but I thought}

\emph{I'd take a look anyway.  With all the averaging out that's going on, maybe}

\emph{the roundoff errors aren't significant anyway.  I've attached the picture to }

\emph{this email.  To me, the picture suggests that $C\sqrt{x}$, [with] $C$ from above, is}

\emph{the right thing to try to prove.  I agree with you, Michael, that a good way}

\emph{to proceed might be by breaking up the sum into two pieces, the 2nd of which}

\emph{can be approximated by the integral.}

:)

$\sim$\emph{ Jared}''

\newpage
\noindent
Here is the picture attached to his email:

\vskip-.05truecm

%\includegraphics [width=\textwidth , bb= 20 20 575 575]{Speck200000terms.jpg} 
% NB: The bb numbers are junk!

\epsfxsize=13cm
\epsfysize=10cm
\centerline{\epsffile{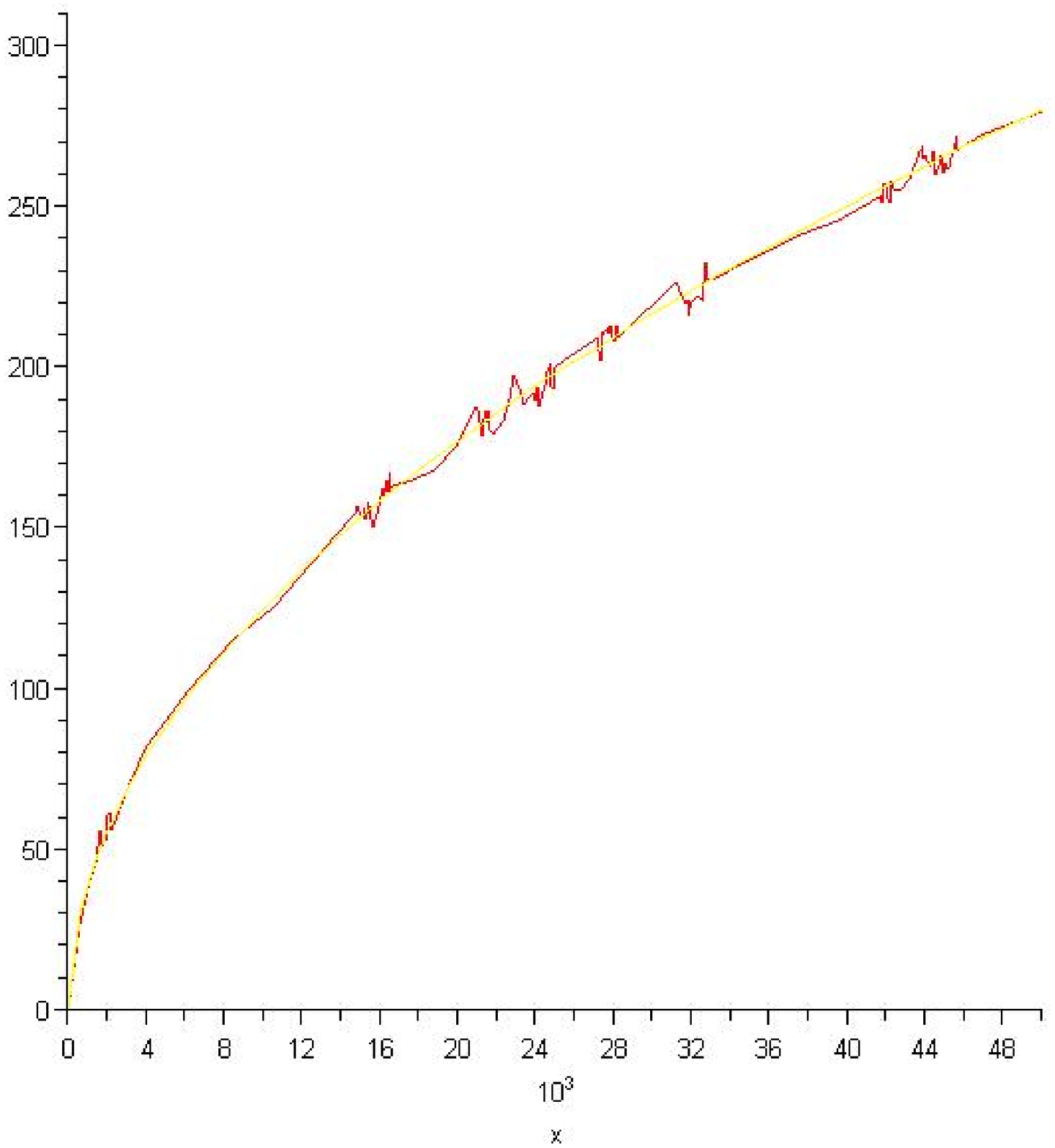}}
\begin{center}
\vskip-3.2truecm
{$\qquad$\scriptsize{Fig.8. (Color online) Speck's 200,000-th partial sum of $\pzcS_2(x)$ together with $\alpha_2^{}\sqrt{x}$.}}
\end{center}

\vskip-.3truecm
\noindent
	Figure 8 is quite remarkable. 
	The agreement of the displayed graph of $\pzcS_2(t)$ with that of $\alpha_2^{}\sqrt{t}$, for
$\alpha_2^{}$ given in \Ref{eq:alphaTWO}, is nothing less than phenomenal.
	The conjecture that  for \emph{all} $t>0$ one has $\pzcS_2(t)=\alpha_2^{}\sqrt{t}+$ ``small'' fluctuations, 
with $\alpha_2^{}$ given in \Ref{eq:alphaTWO}, had to be right!
	
	Incidentally, note that the graph of $\pzcS_2(t)$ shown in Fig.8 displays some intriguing 
intermittency, as known from some turbulent flows. 
	For most $t$ the graph of $\pzcS_2(t)$ is barely distinguishable from that of $\alpha_2^{}\sqrt{t}$, 
but every once in a while an outburst of activity is visible. 
	How fascinating!
	Alas, while preparing for this presentation, when I replotted the graph with much higher 
resolution in Fig.2 the intriguing intermittency disappeared.
	Fig.8 reinforced Jared's intuition about the coefficient $\alpha_2^{}$, which 
would soon be vindicated, but it was quite misleading as a guide for how to think about the fluctuations!

	At noon the next day, $\pzcS_2(t)$ was the topic of the lunch conversation.
	In particular, Mikko Stenlund, at the time postdoc in our mathphys group, 
fell under the spell of the problem.
	Two hours later he sent me the following email:

``\emph{Hi, Michael.}

\emph{For your information, if instead of the series $\sum_n \sin(x/n^2)$ one considers}

\emph{the corresponding integral, one gets with the aid of Fresnel integrals that}

\emph{the asymptotic form is $\sqrt{\pi x/{2}} + \sin(x/\pi^2)-\sin(x)$.}

\emph{Mikko}''

\noindent
	The conjectured $\alpha_2^{}$ integral gives an elementary value for $\alpha_2^{}$ ---
how wonderful!
	Note, though, that the indicated asymptotic expansion (replacing $x\to t$) is for 
$\int_1^\infty\sin(\nu^{-2}t)dt=\sqrt{t}\int_0^t\mu^{-2}\sin(\mu^2)d\mu$, not
the integral at r.h.s.\Ref{eq:SdefZWEIsplitCUBErootINTapprox}.

	Hardly two hours later Jared had completed our proof of the coefficient $\alpha_2^{}$.
	A little upgrading, and also the conjecture about the overall shape of $\pzcS_2(t)$
was proved.
	Our  proof easily generalizes to all $p\!>\!1$, to which I~turn~next.

%%%%%%%%%%%%%%%%%%%%%%%%%%%%%%%%%%%%%%%%%%%%%%%%%%%%%%%%%%%%%%%%%%%%%%%%%%%%%%%%%%%%%%%%%
%%%%%%%%%%%%%%%%%%%%%%%%%%%%%%%%%%%%%%%%%%%%%%%%%%%%%%%%%%%%%%%%%%%%%%%%%%%%%%%%%%%%%%%%%
%%%%%%%%%%%%%%%%%%%%%%%%%%%%%%%%%%%%%%%%%%%%%%%%%%%%%%%%%%%%%%%%%%%%%%%%%%%%%%%%%%%%%%%%%
\subsection{\hskip-.3truecm The overall trend of $\pzcS_p(t)$}
% Proof that $\pzcS_p(t) = \alpha_p^{}{\rm{sign}}(t)|t|^{1/{p}} + O(|t|^{1/(p+1)})\, \forall\, t\in\Rset;\, p>1$
%%%%%%%%%%%%%%%%%%%%%%%%%%%%%%%%%%%%%%%%%%%%%%%%%%%%%%%%%%%%%%%%%%%%%%%%%%%%%%%%%%%%%%%%%
%%%%%%%%%%%%%%%%%%%%%%%%%%%%%%%%%%%%%%%%%%%%%%%%%%%%%%%%%%%%%%%%%%%%%%%%%%%%%%%%%%%%%%%%%
%%%%%%%%%%%%%%%%%%%%%%%%%%%%%%%%%%%%%%%%%%%%%%%%%%%%%%%%%%%%%%%%%%%%%%%%%%%%%%%%%%%%%%%%%
\vskip-.2truecm
	Analogous to the reasoning for when $p=2$, with $t>0$, we now split the summation in the 
series defining $\pzcS_p(t)$ at $n = \lceil{(pt/\tau)^{1/(p+1)}}\rceil =:N_p(t/\tau)$, thus
\begin{equation}
	\pzcS_p(t) 
=\label{eq:SdefSsplitSplusONEroot}
	\textstyle\sum\limits_{n=1}^{N_p(t/\tau)} % {{\lceil{(pt/\tau)^{1/(p+1)}}\rceil}}
		\sin(n^{-p}t)
	+
	\textstyle\sum\limits_{n=N_p(t/\tau)+1}^{\infty}\sin(n^{-p}t). 
%{\lceil{(pt/\tau)^{1/(p+1)}}\rceil}+1}
\end{equation}
	When $\tau$ is small enough (again $\tau < \pi/2$ will do when $t$ gets large), then
for $n > \lceil{(pt/\tau)^{1/(p+1)}}\rceil$ any two consecutive arguments $t/n^p$ and $t/(n+1)^p$ 
of the sine functions will come to lie within one quarter period of sine.
	Moreover, with increasing $n$, for fixed $t/\tau$, the consecutive arguments $t/n^2$ and $t/(n+1)^2$ 
will be more and more closely spaced.
	In other words, for fixed sufficiently small $\tau$, with increasing $t$ 
the part of the sum of $\pzcS_p(t)$ with 
%$n > \lceil{(pt/\tau)^{1/(p+1)}}\rceil$ 
 $n > N_p(t/\tau)$ 
will be an increasingly better Riemann sum approximation of the integral 
%$\int_{\lceil{(pt/\tau)^{1/(p+1)}}\rceil+1}^\infty \sin(\nu^{-p}t)d\nu$.
$\int_{ N_p(t/\tau)+1}^\infty \sin(\nu^{-p}t)d\nu$.
	Thus, and after the variable substitution $\nu^{-p}t=\xi$, 
\begin{equation}
	{\textstyle\sum\limits_{n= N_p(t/\tau)+1}^{\infty}\sin(n^{-p}t)}
\approx \label{eq:SdefSsplitSplusONErootINTapprox}
	t^{1/p}	{\textstyle\frac{1}{p}}
	\int_0^{t/( N_p(t/\tau)+1)^p} \xi^{-1-1/p} \sin \xi d\xi.
\end{equation}
	Since $p>1$, the upper limit of integration at r.h.s.\Ref{eq:SdefSsplitSplusONErootINTapprox} 
goes~to~$\infty$ like $A t^{1/(p+1)}$ when $t\to\infty$, 
and the limiting integral can be evaluated by contour integration:
\begin{equation}
	{\textstyle\frac1p}\int_0^\infty \xi^{-1-1/p} \sin \xi d\xi
= \label{eq:alphaSint}
	\Gamma\big(1-\textstyle\frac1p\big)\sin\big(\textstyle\frac{\pi}{2p}\big).
\end{equation}

\noindent
\emph{Remark}: Integral \Ref{eq:alphaSint} is related by variable substitution to the \emph{generalized Fresnel integral} 
$\int_0^\infty \sin(\eta^q)d\eta = \Gamma\big(1+\textstyle\frac1q\big)\sin\big(\textstyle\frac{\pi}{2q}\big)$, 
which converges for $|q|>1$.

	We will sharpen ``$\approx ... $'' in \Ref{eq:SdefSsplitSplusONErootINTapprox} to ``$= ... + $ 
a subdominant error bound.'' 
	This, a similar estimate comparing r.h.s.\Ref{eq:SdefSsplitSplusONErootINTapprox} 
with $t^{1/p}\times$ r.h.s.\Ref{eq:alphaSint}, and the subdominance of the first sum in \Ref{eq:SdefSsplitSplusONEroot} 
compared to r.h.s.\Ref{eq:SdefSsplitSplusONErootINTapprox}, leads to:

\begin{theorem}
	For all $p>1$, and all $t\in\Rset$, we have
\begin{equation}
	\pzcS_p(t) 
= \label{eq:SsASYMP}
	\alpha_p\, {\rm{sign}}(t)|t|^{1/p}
+
	O\big(|t|^{1/(p+1)}\big),
\end{equation}
with $\alpha_p$ % $=\Gamma\big(1-\textstyle\frac1p\big)\sin\big(\textstyle\frac{\pi}{2p}\big)$
given by r.h.s.\Ref{eq:alphaSint}.
\end{theorem}

\noindent
\emph{Proof:}
	By the anti-symmetry of $\pzcS_p(t)$ it suffices to consider $t>0$, though
we need to distinguish $t\leq t_p$ and $t\geq t_p$ for some $t_p>0$.
	Recall that $\lceil{(pt/\tau)^{1/(p+1)}}\rceil =:N_p(t/\tau)$.
	In all estimates below, $C$ is a generic constant.

	First of all, for $t_p>0$ sufficiently small, we have $\pzcS_p(t)= At +O(t^3)$ for $t\leq t_p$, 
so obviously $|\pzcS_p(t)-\alpha_p^{}t^{1/p}|\leq C t^{1/(p+1)}$ for some $C$ when $t\leq t_p$.

	Turning to $t\geq t_p$, for the first sum at r.h.s.\Ref{eq:SdefSsplitSplusONEroot}
the triangle inequality and then $|\sin \xi|\leq 1$, summing, and an obvious estimate, yield
\begin{equation}
\Big|{\textstyle\sum\limits_{n=1}^{{N_p(t/\tau)}}}\sin(n^{-p}t)\Big|
\quad \leq \quad \label{eq:firstSsumEST}
	\lceil{(pt/\tau)^{1/(p+1)}}\rceil
	\quad \leq \quad C {t^{1/(p+1)}}.
\end{equation}
	For the second sum at r.h.s.\Ref{eq:SdefSsplitSplusONEroot} we find
(for some $\nu_n\in[n,n+1]$)
%\vskip-.6truecm
\begin{eqnarray}
	\Big| {\textstyle\sum\limits_{n={N_p(t/\tau)+1}}^{\infty}\sin(n^{-p}t)}
	- \int_{N_p(t/\tau)+1}^\infty \sin(\nu^{-p}t) d\nu \Big|
&=& \label{eq:secondSsumESTa}\\
	\Big| {\textstyle\sum\limits_{n={N_p(t/\tau)+1}}^{\infty}
	\Big(\sin(n^{-p}t)} - \int_n^{n+1} \sin(\nu^{-p}t) d\nu \Big) \Big|
&=& \label{eq:secondSsumESTb}\\
	\Big| {\textstyle\sum\limits_{n={N_p(t/\tau)+1}}^{\infty}
	\Big(\sin(n^{-p}t)} -  \sin(\nu_n^{-p}t)\Big) \Big|
&=& \label{eq:secondSsumESTc}\\
	\Big|{\textstyle\sum\limits_{n={N_p(t/\tau)+1}}^{\infty}}
	 \int_{t/\nu_n^{p}}^{t/n^{p}} \cos\xi d\xi \Big|
&\leq& \label{eq:secondSsumESTd}\\
	{\textstyle\sum\limits_{n={N_p(t/\tau)+1}}^{\infty}}
	\int_{t/\nu_n^{p}}^{t/n^{p}}	\big|  \cos\xi \big| d\xi 
&\leq& \label{eq:secondSsumESTe}\\
	\textstyle\sum\limits_{n={N_p(t/\tau)+1}}^{\infty}
	t\big(\frac{1}{n^{p}}- \frac{1}{\nu_n^{p}}\big) 
&\leq& \label{eq:secondSsumESTf}\\
	\textstyle\sum\limits_{n={N_p(t/\tau)+1}}^{\infty}
	t\big(\frac{1}{n^{p}}- \frac{1}{(n+1)^{p}}\big) 
&=& \label{eq:secondSsumESTg}\\
	t\lceil{(pt/\tau)^{1/(p+1)}}+1\rceil^{-p}
&\leq& \label{eq:secondSsumESTh}
\!\!	Ct^{1/(p+1)}.
\end{eqnarray}
	In this string,\footnote{I am heeding the advice   
		Michael Fisher gave me (after reading \cite{KiesslingASSISI}) on Dec.14, 2007:\
	``\emph{I have now had a chance to delve further into your write-up. Eventually, I found out 
	why you say {``mean field''}.  The answer is three totally unnumbered 
	equations:  That represents very bad practice!  [...]
		% Please note/recall that I got into the ``business'' of proving a few things rigorously 
		% because of a fatal error of van Hove in an *unnumbered eqn. in the Appendix* of his 
		% claimed proof of the ``everyday'' thermodynamic limit.  
	Please do number all crucial equations in your future papers!}''} 
as for the first three equalities:  \Ref{eq:secondSsumESTa} is manifestly true, whereas
\Ref{eq:secondSsumESTb} holds by the mean value theorem for some $\nu_n\in [n,n+1]$, and 
\Ref{eq:secondSsumESTc} holds by the fundamental theorem of calculus;
as for the ensuing three inequalities: \Ref{eq:secondSsumESTd} holds
by the triangle inequality, \Ref{eq:secondSsumESTe} holds since $|\cos \xi|\leq 1$, 
followed by elementary integration, while \Ref{eq:secondSsumESTf} is due to the monotonic decrease 
of $\nu\mapsto \nu^{-p}$ for $p>1$, with  $\nu_n\in [n,n+1]$;
the ensuing equality \Ref{eq:secondSsumESTg} holds because the sum at l.h.s.\Ref{eq:secondSsumESTg} is telescoping; 
lastly, inequality \Ref{eq:secondSsumESTh} is obvious.

	For the integral in \Ref{eq:secondSsumESTa} the variable substitution $\nu^{-p}t=\xi$ yields
\begin{equation}
	t^{1/p}	{\textstyle\frac{1}{p}}
	\int_0^{t/(N_p(t/\tau)+1)^p}\!\!\!\!\!\! \xi^{-1-1/p} \sin \xi d\xi
= \label{eq:SdefSasympINTrewrite}
	t^{1/p}\Big[\alpha_p^{} -	{\textstyle\frac{1}{p}}
	\int_{t/(N_p(t/\tau)+1)^p}^\infty \!\!\!\!\!\! \xi^{-1-1/p} \sin \xi d\xi\Big].
\end{equation}
	Using one last time the triangle inequality and $|\sin\xi|\leq 1$, we find (for $t\geq 1$):
\begin{equation}
	t^{1/p}{\textstyle\frac{1}{p}} 
	\Big|\int_{t/(N_p(t/\tau)+1)^p}^\infty \xi^{-1-1/p} \sin \xi d\xi\Big|
\ \leq\  \label{eq:SdefSasympINTrewriteESTa}
	 \lceil{(pt/\tau)^{1/(p+1)}+1}\rceil
\ \leq\ C t^{1/(p+1)}.
\end{equation}
%and for $t\leq 1$:
%\begin{equation}
%	t^{1/p}{\textstyle\frac{1}{p}} 	\Big|\int_{t/(N_p(t/\tau)+1)^p}^\infty \xi^{-1-1/p} \sin \xi d\xi\Big|
%\ \leq\  \label{eq:SdefSasympINTrewriteESTb}
%	 \alpha_p^{} t^{1/p}\ \leq\ \alpha_p^{} t^{1/(p+1)}.
%\end{equation}
The entirely elementary proof of Theorem 1 is complete. \qed

\smallskip
	Thm.1 is illustrated below by three graphs of $\pzcS_p(t)$ together with
	their trends $\alpha_p^{} t^{1/p}$, for $p=3/2$, $p=2$, and $p=\surd{7}$.
	The $t$ interval is always $[0,600]$.
	We begin with $p=2$ and $p=\surd{7}$, shown together in Fig.9.

\smallskip

\epsfxsize=10.5cm
\epsfysize=7cm
\centerline{\epsffile{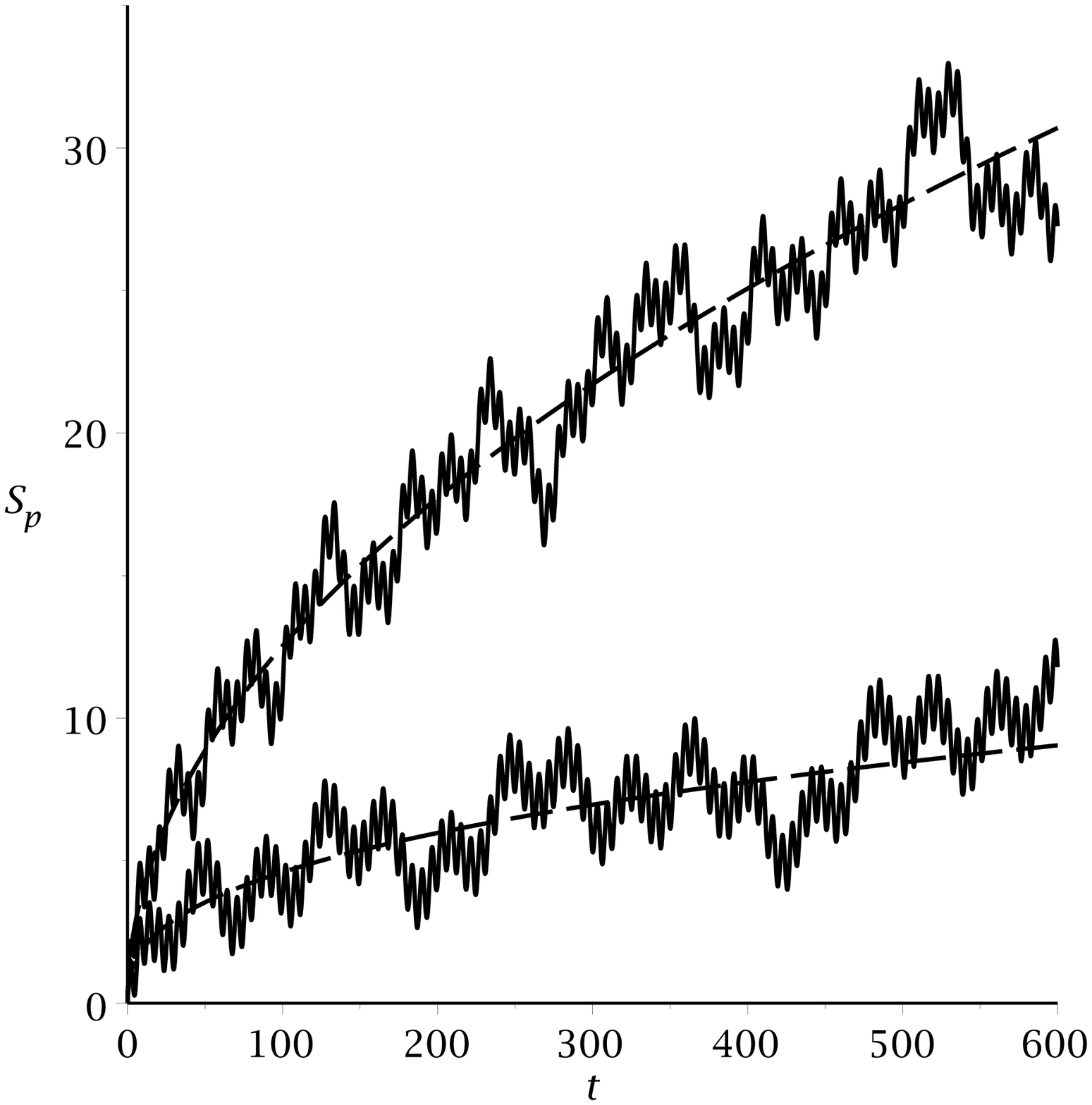}}
\begin{center}
\vskip-.4truecm
{$\qquad$\scriptsize{Fig.9. The 5,000-th partial sums of $\pzcS_p(t)$ for $p=2$ and $p=\surd{7}$,
	together with\\ \hskip1.5truecm their trend functions $\sqrt{\pi t/2}$ and
$\Gamma\big(1-\textstyle\frac{1}{\sqrt{7}}\big)\sin\big(\textstyle\frac{\pi}{2\sqrt{7}}\big) t^{1/\sqrt{7}}$,
respectively.}}
\end{center}

	The case $p=3/2$, shown in Fig.10, is interesting in its own right:
\smallskip

\epsfxsize=10.5cm
\epsfysize=7cm
\centerline{\epsffile{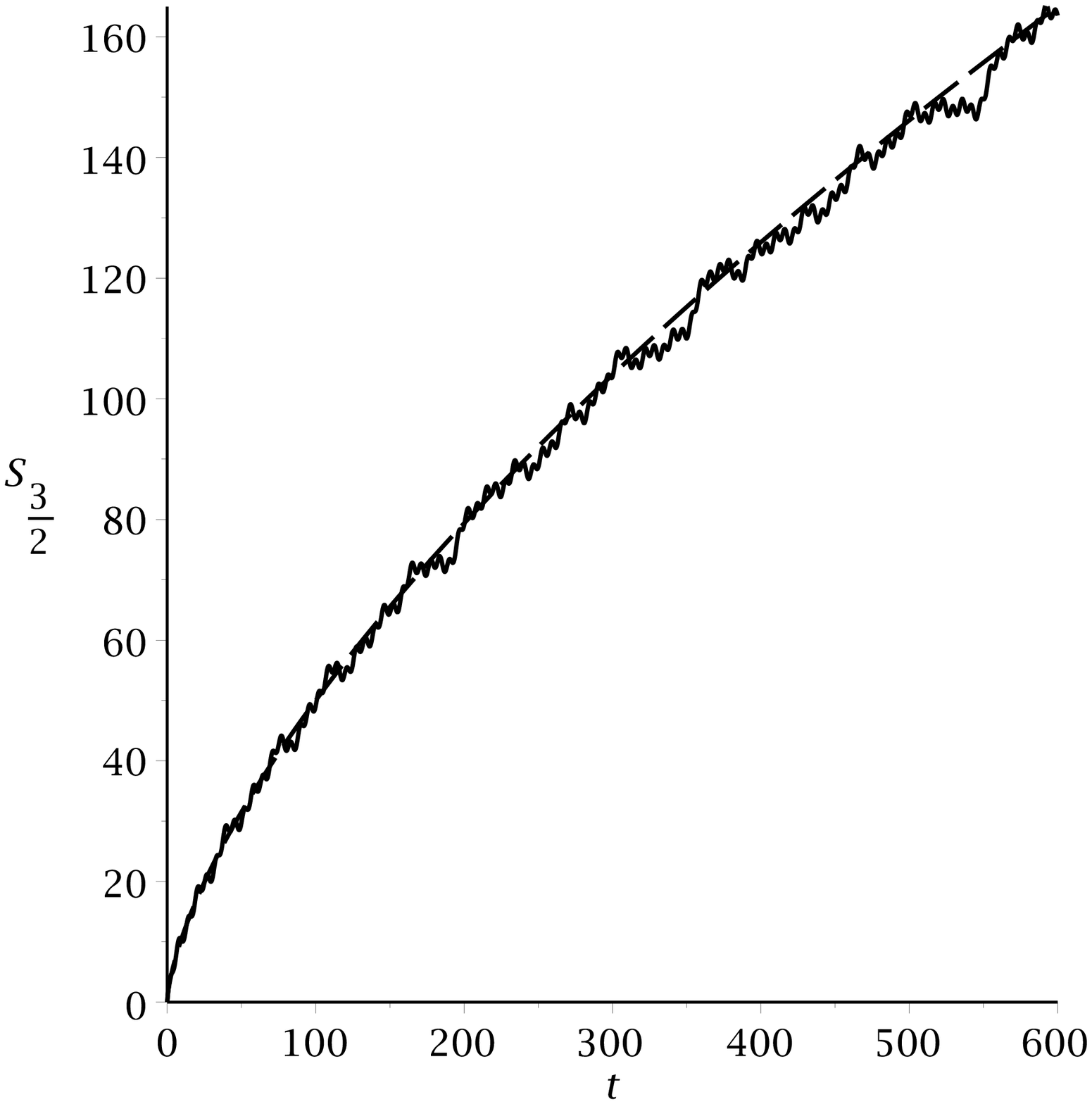}}
\begin{center}
\vskip-.4truecm
{$\qquad$\scriptsize{Fig.10. The 300,000-th partial sum of $\pzcS_{3/2}(t)$ 
	together with $\Gamma\big(\textstyle\frac{1}{3}\big)\sin\big(\textstyle\frac{\pi}{3}\big) t^{2/3}$.}}
\end{center}

\noindent
	Remarkably, a ``staircase'' structure is clearly visible in the graph of $\pzcS_{3/2}(t)$ 
over the $t$-interval $[0,200]$, after which it gets more ``noisy,'' yet for $500<t<550$ another plateau shows.
	Doesn't this call for a number-theoretical explanation?
	
	Since for a moderately small $p$ value like 1.5 a very large number of terms in the partial sum of 
$\pzcS_{3/2}(t)$ was required to achieve a decently converged result, I didn't try to push $p$ close to 1;
except, a mildly smaller, irrational $p=\surd{2}$ was chosen for Fig.6, with a similar expenditure in mode numbers.

	In all three cases shown, the trend function $\alpha_p^{}t^{1/p}$ truly traces the visible trend of $\pzcS_p(t)$.
	The erratic fluctuations about the trend are more slowly growing in amplitude than the trend. 
	Our Thm.1 says that they are bounded in amplitude by $O(t^{1/(p+1)})$. 
	To get an idea of how accurate this bound is, I resorted to Maple to plot  
$\pzcS_p(t)$ - $\alpha_p^{} t^{1/p}=:\triangle \pzcS_p(t)$ together with $\pm \beta_pt^{1/(p+1)}$ for 
$p=2$ and  $p=\surd{7}$, with empirically near-optimized $\beta_p$, see Figs. 11 and 12:

\epsfxsize=10.5cm
\epsfysize=7cm
\centerline{\epsffile{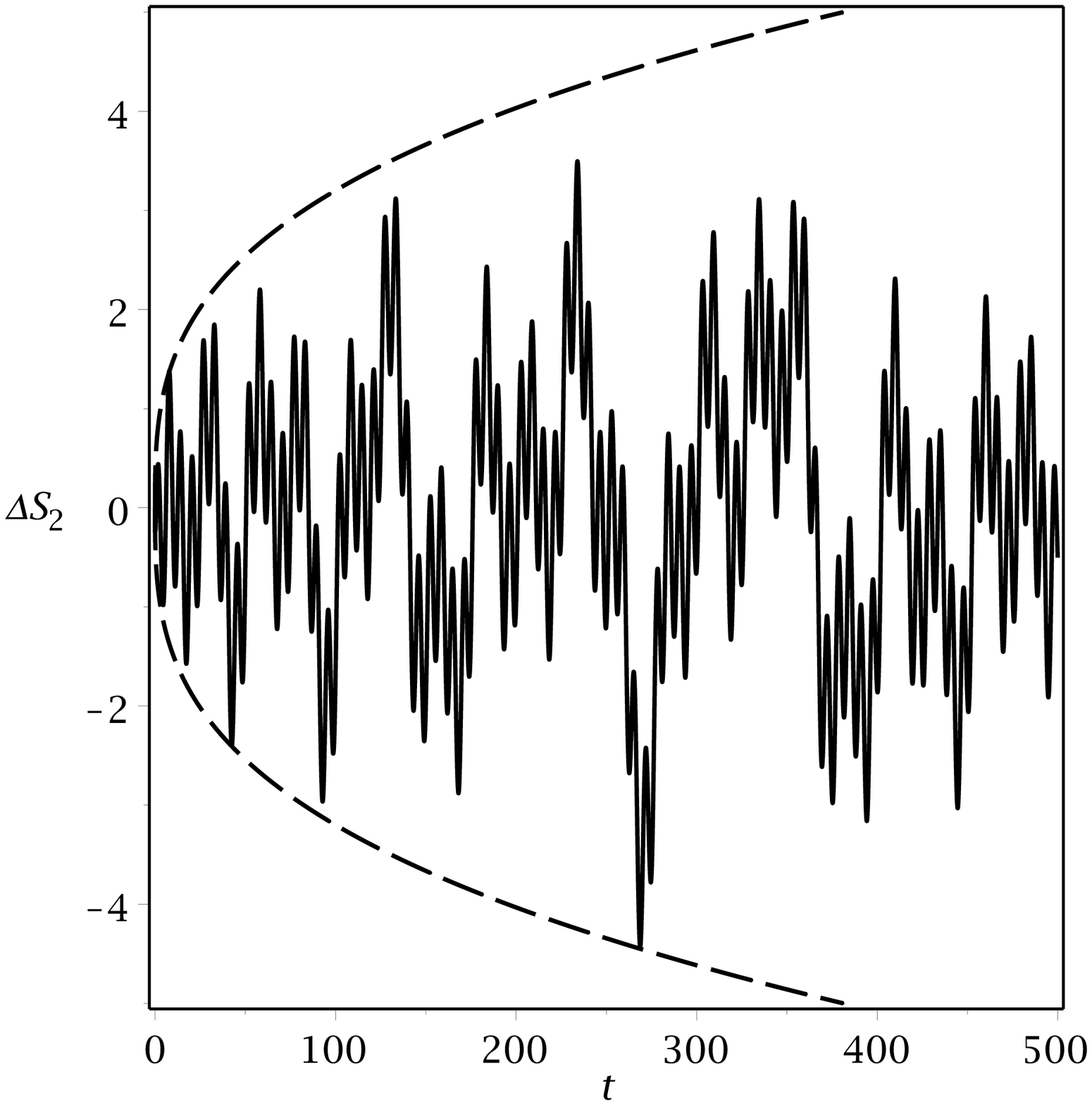}}
\begin{center}
\vskip-.6truecm
{$\qquad$\scriptsize{Fig.11. The 200,000-th partial sum of $\pzcS_2(t)$ - $\sqrt{\pi t/2}$
	together with $\pm \frac{20}{29}t^{1/3}$.}}
\end{center}

\vskip-.3truecm

\epsfxsize=10.5cm
\epsfysize=7cm
\centerline{\epsffile{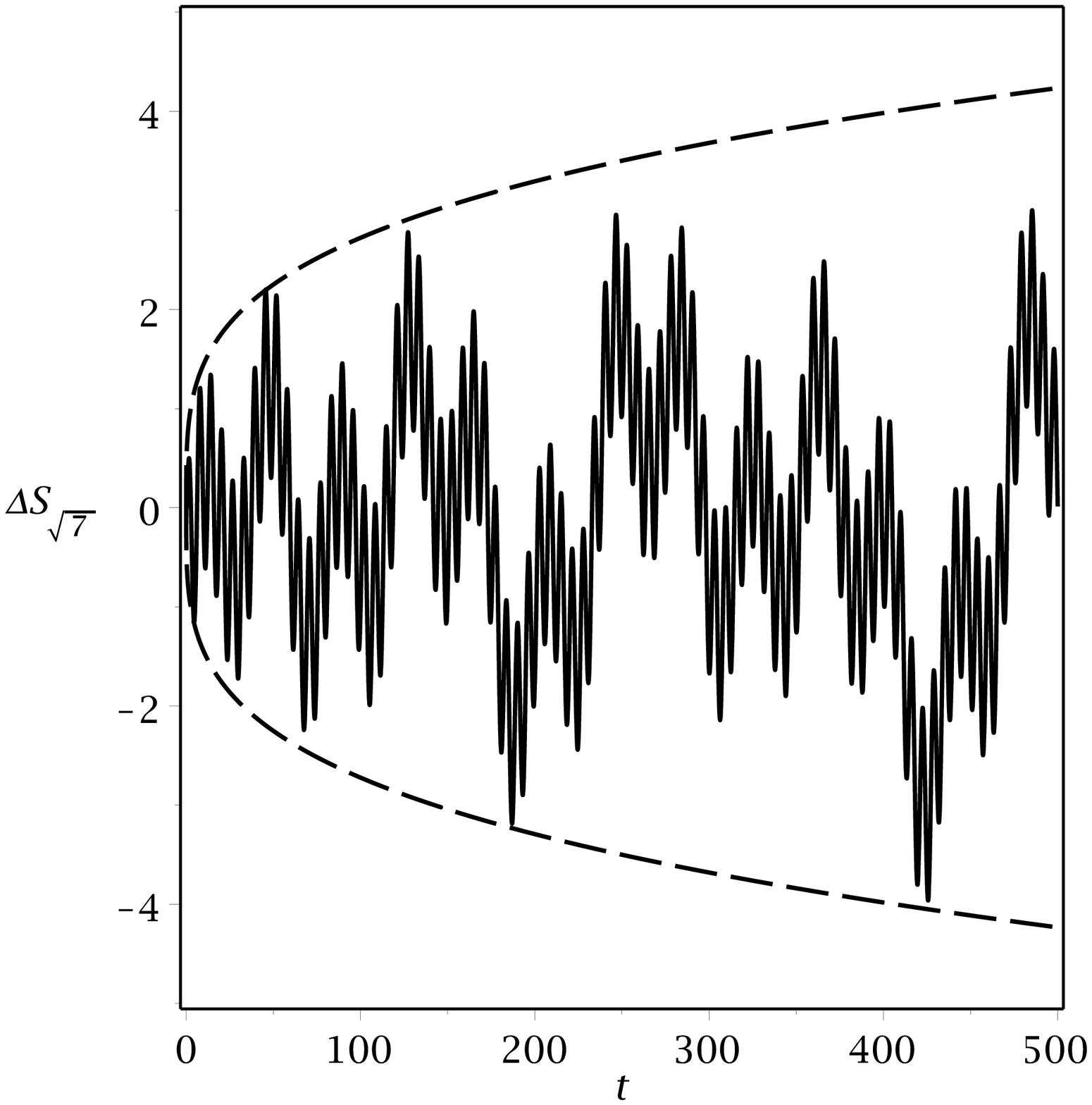}}
\begin{center}
\vskip-.6truecm
{$\qquad$\scriptsize{Fig.12. The 20,000-th partial sum of $\pzcS_{\surd{7}}(t)- 
\Gamma\big(1-\textstyle\frac{1}{\surd{7}}\big)\sin\big(\textstyle\frac{\pi}{2\surd{7}}\big) t^{1/\sqrt{7}}$
	with $\pm 0.77t^{1/(1+\surd{7})}$.}}
\end{center}

\noindent
	Figs.11 and 12 reveal that the functions $\pm\beta_pt^{1/(p+1)}$ are accurately bounding the growth of 
the largest fluctuation amplitudes over the shown $t$ interval with empirically optimized $\beta_p$;
I have not tried analytically to optimize $\beta_p$.
	Of course, a larger sample of $p$ values would allow a more reliable conclusion; however, 
since much higher precision was needed for these figures, I plotted only the cases $p=2$ and $p=\surd{7}$.
	In section 4.2 we will use a change of variables which allows us to plot $\triangle\pzcS_p$ for larger
$t$ values, see Figs. 14 \&\ 15.

	After this three-day flurry of activity the inquiry into $\pzcS_p(t)$ stopped almost as abruptly 
as it had started, because all participants had to return to their own important businesses.
	However, pre-conditioned by my upbringing in statistical mechanics, I resolved to resume the 
inquiry into the fluctuations of $\pzcS_p(t)-\alpha_p^{}t^{1/p}$ whenever the opportunity would arise.

%%%%%%%%%%%%%%%%%%%%%%%%%%%%%%%%%%%%%%%%%%%%%%%%%%%%%%%%%%%%%%%%%%%%%%%%%%%%%%%%%%%%%%%%%
%%%%%%%%%%%%%%%%%%%%%%%%%%%%%%%%%%%%%%%%%%%%%%%%%%%%%%%%%%%%%%%%%%%%%%%%%%%%%%%%%%%%%%%%%
%%%%%%%%%%%%%%%%%%%%%%%%%%%%%%%%%%%%%%%%%%%%%%%%%%%%%%%%%%%%%%%%%%%%%%%%%%%%%%%%%%%%%%%%%
%%%%%%%%%%%%%%%%%%%%%%%%%%%%%%%%%%%%%%%%%%%%%%%%%%%%%%%%%%%%%%%%%%%%%%%%%%%%%%%%%%%%%%%%%
\section{Statistics of the fluctuations of $\pzcS_p(t)-\alpha_p^{}t^{1/p}$}
%%%%%%%%%%%%%%%%%%%%%%%%%%%%%%%%%%%%%%%%%%%%%%%%%%%%%%%%%%%%%%%%%%%%%%%%%%%%%%%%%%%%%%%%%
%%%%%%%%%%%%%%%%%%%%%%%%%%%%%%%%%%%%%%%%%%%%%%%%%%%%%%%%%%%%%%%%%%%%%%%%%%%%%%%%%%%%%%%%%
%%%%%%%%%%%%%%%%%%%%%%%%%%%%%%%%%%%%%%%%%%%%%%%%%%%%%%%%%%%%%%%%%%%%%%%%%%%%%%%%%%%%%%%%%
%%%%%%%%%%%%%%%%%%%%%%%%%%%%%%%%%%%%%%%%%%%%%%%%%%%%%%%%%%%%%%%%%%%%%%%%%%%%%%%%%%%%%%%%%
%
%%%%%%%%%%%%%%%%%%%%%%%%%%%%%%%%%%%%%%%%%%%%%%%%%%%%%%%%%%%%%%%%%%%%%%%%%%%%%%%%%%%%%%%%%
%%%%%%%%%%%%%%%%%%%%%%%%%%%%%%%%%%%%%%%%%%%%%%%%%%%%%%%%%%%%%%%%%%%%%%%%%%%%%%%%%%%%%%%%%
%%%%%%%%%%%%%%%%%%%%%%%%%%%%%%%%%%%%%%%%%%%%%%%%%%%%%%%%%%%%%%%%%%%%%%%%%%%%%%%%%%%%%%%%%
\subsection{Kac's central limit theorem}
%{Andrei Markov and Marek Kac}
%%%%%%%%%%%%%%%%%%%%%%%%%%%%%%%%%%%%%%%%%%%%%%%%%%%%%%%%%%%%%%%%%%%%%%%%%%%%%%%%%%%%%%%%%
%%%%%%%%%%%%%%%%%%%%%%%%%%%%%%%%%%%%%%%%%%%%%%%%%%%%%%%%%%%%%%%%%%%%%%%%%%%%%%%%%%%%%%%%%
%%%%%%%%%%%%%%%%%%%%%%%%%%%%%%%%%%%%%%%%%%%%%%%%%%%%%%%%%%%%%%%%%%%%%%%%%%%%%%%%%%%%%%%%%
	Years later, in March 2011, while listening to Felix Izrailev's interesting 
presentation about quantum thermalization at Michael Kastner's {\sc{sti$\alpha$s}} workshop in Stellenbosch,
South Africa, I noticed that he refered to work by Mark Kac on the central limit theorem for certain 
trigonometric series.
	I immediately wondered whether this was the information I had been waiting for to hear! 
	
	The relevant original publications are \cite{KacFRENCH} and \cite{KacAJM}, which together with
some other works by Kac were expanded into his book \cite{KacBOOK}.
	According to the charming Kac memoir by Henry McKean (cf. p.219 in \cite{McKean})\footnote{Note
		a typo $1/2$, instead of $1/\surd{2}$, in the pertinent formula on p.219 in \cite{McKean}.}, 
in \cite{KacFRENCH} and \cite{KacAJM} the following is proved:
\begin{theorem}
	Let the set of frequencies $\{\omega_n\}_{n\in\Nset}$ be linearly independent over $\Qset$ 
(i.e., for any $N\in\Nset$, the only solution to $\sum_{n=1}^N z_n\omega_n =0$ with all $z_n\in\Zset$ is
$z_1=\cdots=z_N=0$).
	Let ``$\meas$'' denote Lebesgue measure on $\Rset$.
	Then 
\begin{equation}
\hskip-3pt
	\lim_{N\to\infty}
	\lim_{T\to\infty}
	{\textstyle\frac{1}{T}}
	\meas\Big\{t\in[0,T]: a\leq {\textstyle\sqrt{\frac{2}{N}}\sum\limits_{n=1}^N}\sin \omega_nt\leq b\Big\}
= \label{eq:KacCLT}
	{\textstyle\frac{\,1}{\sqrt{2\pi}}}\!\int_a^b\! e^{-\frac{1}{2}y^2}dy.
\end{equation}
\end{theorem}

	Several remarks are in order: in \cite{KacAJM}, and in more detail again in \cite{KacBOOK},
Kac himself derives a similar formula in which cosines replace the sine functions, and with the $t$-average 
taken over the interval $[-T,T]$ rather than $[0,T]$.
	For Kac's cosine theorem it is obvious (since cosine is an even function) that the average over $[-T,T]$ equals
the one over $[0,T]$; however, this is not true for a sum of odd sine functions.
	One has to go through Kac's cosines proof to see that, after replacing $[-T,T]$ with $[0,T]$ averages, 
one can also work with sine replacing cosine, indeed.

	At first glance Kac's Theorem looks just like ``what the doctor ordered'' for $\pzcS_p(t)$.
	Unfortunately, what it says about ``$\pzcS_p(t)$'' (for suitable $p$)
is \emph{not} about $\pzcS_p(t)$ --- instead, it is about the infinite time averages of the family 
of partial sums of $\pzcS_p(t)$.
	Of course, $\pzcS_p(t)$ is \emph{defined} as the limit $N\to\infty$ of the sequence of its $N$-th partial sums,
but this limit does not commute with the limit $T\to\infty$ of time-averages over intervals of length $T$;
only for a fixed partial sum, $t$-averaging over $[0,T]$ and summation do commute.
	Kac's theorem demands that for any $N$-th partial sum of $\pzcS_p(t)$ one first performs the limit~$T~\to~\infty$ 
for the average amount of time this partial sum eventually spends in the interval $[a\sqrt{N/2},b\sqrt{N/2}]$, then lets 
$N\to\infty$ (cf. \cite{KacAJM}, \cite{KacBOOK}).

	Kac's theorem implements Steinhaus' notion of ``statistical independence of functions:'' 
the average amounts of time which individual sine functions with incommensurate frequencies 
spend in any infinitesimal interval within $[-1,1]$ are eventually \emph{i.i.d.} random 
variables with mean zero and standard deviation $1/\surd{2}$ --- thus the central limit theorem 
type appearance of his theorem.
	Recall that the limit $T\to\infty$ of the unrestricted $t$-average over $[0,T]$ of \emph{each} sine function 
vanishes whereas the $t$-average of its square converges to 1/2.

	By contrast, we need to take a time average \emph{after} the infinite summation over all sine functions
has been carried out and the trend function subtracted. 
	This makes it plain that Theorem 2 above is not applicable to our problem!
	
	Now, all this does not mean that the fluctuations of $\pzcS_p(t) -\alpha_p^{}t^{1/{p}}$ are not normal
--- they may well be (``Not so!'' is the comment of one of the referees --- see below).
	At the time of my SMM~106 presentation, under the spell of Kac's central limit theorem, I indeed 
conjectured that, after ``suitable $p$-dependent rescaling,'' a normal law should hold for the fluctuations of 
$\pzcS_p(t) -\alpha_p^{}t^{1/{p}}$, at least for irrational $p$ (NB: As explained in the introduction, another
referee noted that $p\not\in\Qset$ won't be sufficient to guarantee the rational linear independence of the involved
frequencies  $\{n^{-p}\}_{n\in\Nset}$.) 
	More precisely, a careful inspection of Kac's proof, which is based on L\'evy's rigorous version of 
Markov's method of characteristic functions (i.e., Fourier transforms of probability measures), reveals that 
Markov's method should also determine the distribution of the fluctuating values of 
$\pzcS_p(t) -\alpha_p^{}t^{1/{p}}$, yet it also is clear that from some point on Kac's arguments will 
have to be modified.
	More to the point, even though our theoretical error bounds are too rough to show it, empirically the second
term at r.h.s.\Ref{eq:SdefSsplitSplusONEroot} seems to be a very accurate Riemann sum approximation for the trend function
when $t/\tau$ becomes moderately large, which means that $\triangle\pzcS_p(t)$ should be well approximated by the first
term at r.h.s.\Ref{eq:SdefSsplitSplusONEroot}, which is a finite sum at each $t$, containing not more than
$N_p(T/\tau)$ terms in the $t$-averages over $[0,T]$.
	Since $N_p(T/\tau)\asymp C T^{1/(p+1)}$, the longest wavelength in the partial
sum of sines, which is averaged over $[0,T]$, grows basically $\propto T^{p/(p+1)}$, i.e. sublinear in $T$
so that, as $T$ grows large, even the sine functions with the longest wavelengths in the partial sum are averaged over 
 many cycles, infinitely many in the limit $T\to\infty$. 
	The upshot is the following conjecture (extending Kac's ``central limit theorem''):
\vskip-.4truecm
\begin{conjecture}
	Suppose $p$ is chosen so that the frequencies $\{n^{-p}\}_{n\in\Nset}$ are rationally linear independent.
	Then

\vskip-.6truecm
\begin{equation}
\hskip-.4truecm
	\lim_{T\to\infty}
	{\textstyle\frac{1}{T}}
	\meas\Big\{t\in[0,T]\!: a\leq {\textstyle\sqrt{\frac{2}{N_p(\frac{t}{\tau})}}
	\!\!\sum\limits_{n=1}^{\ N_p(\frac{t}{\tau})}\!\!\sin(n^{-p}t)}\leq b\Big\}
= \label{eq:SpCLTconj}
	{\textstyle\frac{\,1}{\sqrt{2\pi}}}\!\int_a^b\!\! e^{-\frac{1}{2}y^2}\!dy.
\end{equation}
\end{conjecture}
\vskip-.1truecm

	Note that Conjecture 1 is weaker than conjecturing that the fluctuations of 
$\pzcS_p(t) -\alpha_p^{}t^{1/{p}}=\triangle\pzcS_p(t)$ itself are eventually normal.
	In any event, Conjecture 1 is interesting in its own right and, if true, may serve as 
an important stepping stone on the way to characterizing the fluctuations of $\triangle\pzcS_p(t)$.
	I hope to settle this issue at some point, or that someone else will feel inspired to do so.

%%%%%%%%%%%%%%%%%%%%%%%%%%%%%%%%%%%%%%%%%%%%%%%%%%%%%%%%%%%%%%%%%%%%%%%%%%%%%%%%%%%%%%%%%
%%%%%%%%%%%%%%%%%%%%%%%%%%%%%%%%%%%%%%%%%%%%%%%%%%%%%%%%%%%%%%%%%%%%%%%%%%%%%%%%%%%%%%%%%
\subsubsection{Experts' opinions on the fluctuations of $\pzcS_p(t) -\alpha_p^{}t^{1/{p}}$}
%%%%%%%%%%%%%%%%%%%%%%%%%%%%%%%%%%%%%%%%%%%%%%%%%%%%%%%%%%%%%%%%%%%%%%%%%%%%%%%%%%%%%%%%%
%%%%%%%%%%%%%%%%%%%%%%%%%%%%%%%%%%%%%%%%%%%%%%%%%%%%%%%%%%%%%%%%%%%%%%%%%%%%%%%%%%%%%%%%%

	All three referees confirmed my hunch that the questions raised above, and many more, can be answered with the
techniques of analytical number theory, see the books \cite{Vinogradov}, \cite{expSUMSbook}, and the surveys
\cite{Montgomery}, \cite{IwaKowa}.
	In particular, one of the referees noted: ``Assuming the Riemann hypothesis, which is quite normal 
in such type [of] studies, I was able to show that for any $\veps>0$,
\begin{equation}
	\pzcS_p(t) 
= \label{eq:SpTRENDplusERROR}
	\alpha_p^{}t^{1/{p}} + t^{1/2(p+1)+\veps} g_{p,\veps}(\ln t),
\end{equation}
where $g_{p,\veps}$ is an $L^2$-function, so that $\int_{-\infty}^\infty |g_{p,\veps}(u)|^2 du < \infty$.
	I think that this gives a much stronger control of the error term than [\Ref{eq:SsASYMP}].''
	Another referee made a similar observation about $\pzcS_p(t)$ as expressed in \Ref{eq:SpTRENDplusERROR},
namely that by standard techniques of Fourier analysis and analytic number theory as
explained, e.g., in \cite{expSUMSbook}, one would be able to prove that
\begin{equation}
	\pzcS_p(t) 
= \label{eq:expSUMcontrolSp}
	\alpha_p^{}t^{1/{p}}	+ t^{1/2(p+1)} E_p(t);\qquad t>1,
\end{equation}
where $E_p(t)= O(t^{o(1)})$ (for $t>1$).
	This referee further noted that ``at least in some cases (like $p=2$) it seems possible
to use the methods from the paper \cite{HeathBrown} to understand the distribution of $E_p(t)$. 
	In any case, I think it is easy to show that for any $p$ the distribution [of $E_p(t)$]
is not going to be normal ... % (because the small $j$ are going to contribute the most) 
which would show that this conjecture [about the fluctuations of $\triangle\pzcS_p(t)$] made by the author is wrong.''
	(The referee also outlined the exponential sum obtained for $E_p(t)$ with
the methods in \cite{expSUMSbook}, and why the distribution of $E_p(t)$ should be non-normal.
	Unfortunately, since I am no expert in analytical number theory, I refrain from making
an (inevitably amateurish) attempt to explain these arguments here.)
	The third referee similarly noted that ``the Gaussian behavior is violated in 
other examples in number theory under a similar philosophy \cite{BleherETal} (see also \cite{Bleher}).''

	I am grateful to stand rectified about my speculations about the normality of the fluctuations of 
$\triangle\pzcS_p(t)$.
	I am also grateful for the observation that $p\not\in\Qset$ is insufficient to guarantee rational linear
independency of the set of frequencies $\{n^{-p}\}$.
	As far as I can see, though, my (so revised) Conjecture 1 about the normality of fluctuations of 
the $\lceil t^{1/(p+1)}\rceil$-th partial sum of $\pzcS_p(t)$ is still viable.

%%%%%%%%%%%%%%%%%%%%%%%%%%%%%%%%%%%%%%%%%%%%%%%%%%%%%%%%%%%%%%%%%%%%%%%%%%%%%%%%%%%%%%%%%
%%%%%%%%%%%%%%%%%%%%%%%%%%%%%%%%%%%%%%%%%%%%%%%%%%%%%%%%%%%%%%%%%%%%%%%%%%%%%%%%%%%%%%%%%
%%%%%%%%%%%%%%%%%%%%%%%%%%%%%%%%%%%%%%%%%%%%%%%%%%%%%%%%%%%%%%%%%%%%%%%%%%%%%%%%%%%%%%%%%
\subsection{Late-$t$ asymptotics of $\pzcS_p(t)$: Riemann's $\zeta$ function}
%%%%%%%%%%%%%%%%%%%%%%%%%%%%%%%%%%%%%%%%%%%%%%%%%%%%%%%%%%%%%%%%%%%%%%%%%%%%%%%%%%%%%%%%%
%%%%%%%%%%%%%%%%%%%%%%%%%%%%%%%%%%%%%%%%%%%%%%%%%%%%%%%%%%%%%%%%%%%%%%%%%%%%%%%%%%%%%%%%%
%%%%%%%%%%%%%%%%%%%%%%%%%%%%%%%%%%%%%%%%%%%%%%%%%%%%%%%%%%%%%%%%%%%%%%%%%%%%%%%%%%%%%%%%%
\vskip-.2truecm
	Sometimes wondrous things happen.
	After reading the announcements of the SMM~106 conference talks,
Norm Frankel and Steve Miller requested the pdf file of my upcoming talk.
	Since, as usual, I finished the preparations for my talk barely in time, they had 
to wait until then.
	Soon after, they got back to me with exciting emails, the essence of each 
of which I pool together. 
	Here first is Norm Frankel (whose inimitable style I like to preserve):
\smallskip

``\emph{Dear Michael,}

\emph{MILLE GRAZIE! }

\emph{I've been looking forward to receiving this and will read and study it with}

\emph{relish [and mustard - HI]. I'll write back when I have $> \ln(2)$ to say. $\dots$}

%\centerline{$\vdots$}

% \emph{THE HAPPIEST OF HOLIDAYS FOR YOU AND YOURS!}
% \emph{Warmest Regards, Norm}''
% ``Dear Michael,

\emph{Using a Mellin transform, the asymptotics comes out in one line. I find}

$\pzcS_2(x) \sim (1/4) \sqrt{2\pi/x} - \pi/4 +$\emph{intricate terms. Similarly $\pzcS_p(x)$ can be readily}

%NB: I corrected Norm's 2pi/x into 2pi x

\emph{exhibited. I seem to be differing by a factor of $1/2$. $\dots$}

% Cheers, Norm''
%``Dear Michael,
%\centerline{$\vdots$}

\emph{The results I sent were for $\pzcS_{-2}(x)$, not $\pzcS_2(x)$. OF COURSE yours is a}

\emph{MUCH trickier sum --- back 'gain soon. $\dots$}

%Cheers, Norm''

%\centerline{$\vdots$}

%``Dear Michael,
%\emph{I enjoy your slides. Mille grazie!}

\emph{I've just looked again at your sum. I think what I started out to do is}

\emph{correct in concert with analytical continuation: the Mellin transform of}

$\sin(n^p x) = \sin(\pi s/2) \Gamma[s] \zeta(ps) x^{-s}$. \emph{Inverting readily gives the large $x$}

\emph{asymptotics --- even for $p=-2$, your series --- correction terms follow.}

\emph{It may be incorrect; if so, mia culps. $\dots$}

\emph{THE HAPPIEST OF HOLIDAYS FOR YOU AND YOURS!}

\emph{Warmest Regards, Norm}''

%Happy Holidays, Norm''
%
% NB: I used the ending of the first email as ending of the pooled 
%

\noindent
And here is Steve Miller:

``\emph{Dear Michael,}

\emph{thank you very much for your e-mail. You really made excellent powerpoint}

\emph{slides\! --\! I am sure the audience appreciated them. They were very clear and}

\emph{entertaining. [...]}\footnote{Steve drew my attention to the intriguing papers \cite{ChUb, MillerSchmid}
		on $\sin(n^2x)$ series.}
%{Thanks also for the ``plug'', which we appreciate.}
%\emph{I should have also mentioned that Iwaniec had a student, Fernando Chamizo,}
%\emph{who wrote about trigonometric series of the Riemann-Weierstrass type,}
%\emph{but with $n^{positive\ power}$ in the argument.}
\emph{I had not seen anything with $n^{-2}$ like you have.}
%\centerline{}

\emph{I have a few comments which I hope will be helpful to you. You asked about}

\emph{the connection with the Riemann $\zeta$ function at the end of your talk. If you}

\emph{(formally,\! at least)\! take\! the Mellin transform of the function $\!f\!\!:\!y\! \mapsto\! \sin(x/y^2)$}

\emph{which is $Mf(s) = -{\frac12} x^{s/2} \Gamma(-s/2) \sin(\pi s/4)$, then your sum is a contour}

\emph{integral of $Mf(s)$ times the Riemann $\zeta$ function. The Riemann $\zeta$ function}

\emph{has only one pole, at $s=1$,\! and the residue of $Mf(s)\zeta(s)$ at $s=1$ is your}

\emph{main term $\sqrt{\pi x/2}$. The only other poles of $Mf(s)\zeta(s)$ are at values of $s$}

\emph{of the form $2+4n,\ n \geq 0$; these give correction terms in the asymptotic}

\emph{expansion of Greenfield's infinite sum as $x\to 0$. So the full asymptotic}

\emph{expansion should come from this.}

\emph{I also want to note that the functional equation of the Riemann $\zeta$ function}

\emph{(again, completely formally -- this is essentially Poisson summation here)}

\emph{gives the identity that your Greenfield's sum is the sum of $g(n,x)$ over} 

\emph{$n>0$, where the Mellin transform of $g(z,x)$ in $z$ is}

\centerline{$-(2 \pi)^{1-s} x^{(1-s)/2} \cos(\pi(s+1)/4) \Gamma(s-1) \zeta(s)/\Gamma((1-s)/2)$}

\emph{This function (without the $\zeta(s)$ factor) is a sum of hypergeometric functions}

\epsfxsize=9cm
\epsfysize=4cm
\centerline{\epsffile{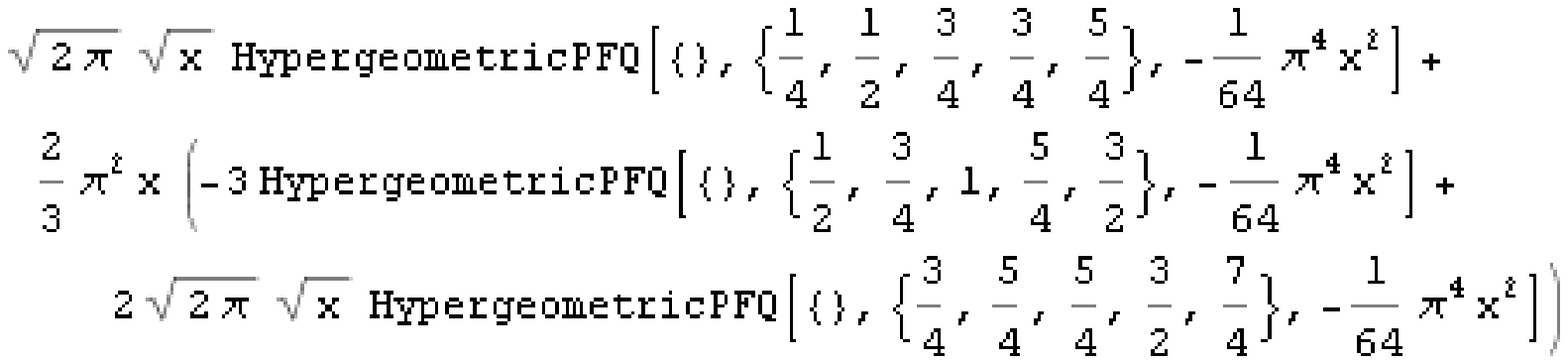}}
\begin{center}
\vskip-1truecm
{\scriptsize{Fig.13. The sum of hypergeometric functions Steve included as image.}} %image.png 
\end{center}

\vskip-10pt
\emph{Mellin inversion shows Greenfield's sum is also a sum of this function.}

\emph{That dual point of view may reveal some other properties of the sums you}

\emph{considered in your slides. [I think I botched a calculation here,
	but I hope}

\emph{the idea and strategy was clear].}

\emph{Best holiday wishes,}

\emph{Steve}''

	I was thrilled --- and immediately turned contemplative:
	While Norm is located half around the globe away from my office, Steve Miller's office isn't much further 
away from mine than Steve Greenfield's! 
	Here I was, having the problem in the back of my mind all these years --- without ever mentioning it to 
Steve (M.)?
	What if Steve G. would have sent his question to Steve~M. instead? 
	It reminded me of ``Missed Opportunities'' \cite{Dyson}, the beautiful Gibbs lecture by 
Norm Frankel's longtime friend Freeman Dyson.

	Enlightened by their comments I decided to compute the late time asymptotics of $\pzcS_p(t)$ beyond
the leading order term.
	However, things aren't quite as straightforward as they seem!

	First, let me flesh out what Norm and Steve wrote in their emails.
	Recall that the \emph{Mellin transform} of a continuous function $f:\Rset_+\to \Rset$ is %defined by
\begin{equation}
	(\cM f)(s) 
:= \label{eq:MellinTdef}
	\int_0^\infty y^{s-1}f(y)dy
\end{equation}
wherever r.h.s.\Ref{eq:MellinTdef} is well-defined. 
	Supposing $|f(y)|=O(y^{-a})$ for $y\downarrow 0$ and $|f(y)|=O(y^{-b})$ for $y\uparrow\infty$,
with $a<b$ sharp, then $(\cM f)(s)$ is analytic in its fundamental strip $a\!<\Re s <\!b$,
where it tends to zero as $|\Im(s)|\to\infty$ (by the Riemann--Lebesgue lemma).
	Writing $(\cM f)(s)=:\tilde{f}(s)$, its inverse transform is given by
the straight contour integral (in the improper Riemann sense)
\begin{equation}
	f(y) 
=  \label{eq:MellinTinv}
	{\textstyle\frac{1}{2\pi i}}\int_{c-i\infty}^{c+i\infty} y^{-s}\tilde{f}(s)ds
\qquad \mathrm{any}\ c\in(a,b).
\end{equation}

	Turning to $\pzcS_p(t)$, for $t>0$  we define $r:=t^{-1/p}$ and write $\pzcS_p(t)=\pzcR_p(r)$, i.e.
\begin{equation}
	\pzcR_p(r) 
=\label{eq:Rdef}
	\textstyle\sum_{n\in\Nset}^{}\sin\big((nr)^{-p}\big); \qquad  p>1.
\end{equation}
	Since $|\pzcR_p(r)|= O(r^{-1})$ for $r\downarrow 0$ and $|\pzcR_p(r)|= O(r^{-p})$ for $r\uparrow\infty$,
for $1<\Re s< p$ we can take the Mellin transform of $\pzcR_p(r)$ to find, with obvious manipulations, 
\begin{eqnarray}
	\widetilde\pzcR_p(s) 
&=&\label{eq:MellinRa}
	\int_0^\infty r^{s-1}\textstyle\sum\limits_{n\in\Nset}^{} \sin\big((nr)^{-p}\big)dr\\
&=&\label{eq:MellinRb}
	{\textstyle\sum\limits_{n\in\Nset}^{}}\int_0^\infty r^{s-1}\sin\big((nr)^{-p}\big)dr\\
&=&\label{eq:MellinRc}
	{\textstyle\sum\limits_{n\in\Nset}^{}\frac{1}{n^s}}\int_0^\infty y^{s-1}\sin(y^{-p})dy\\
&=&\label{eq:MellinRd}
	\zeta(s)
	\textstyle\frac{1}{p}\Gamma\big(\!-\frac{s}{p}\big)\sin\!\big(\!-\frac{\pi}{\scriptstyle{2}}\frac{s}{p}\big).
\end{eqnarray}
which is analytic in $1 < \Re s<p$, its fundamental strip.
	And so, for $c\in(1,p)$,
\begin{equation}
	\pzcR_p(r) 
=\label{eq:RpMELLINrep}
	{\textstyle\frac{1}{2\pi i}}\int_{c-i\infty}^{c+i\infty} r^{-s}
     \zeta(s)
\textstyle\frac{1}{p}\Gamma\big(\!-\frac{s}{p}\big)\sin\!\big(\!-\frac{\pi}{\scriptstyle{2}}\frac{s}{p}\big)ds,
\end{equation}
or, after switching back to $t= r^{-p}$, renaming $s/p$ into $\varsigma$, and introducing
\begin{equation}
	\bar\pzcQ_p(\varsigma) 
=\label{eq:QpMELLINdef}
	\zeta(p\varsigma)
	\Gamma(-\varsigma)\sin\!\big(\!-\textstyle{\frac{\pi}{\scriptstyle{2}}}\varsigma\big),
\end{equation}
we find
\begin{equation}
	\pzcS_p(t) 
=\label{eq:SpMELLINrep}
	{\textstyle\frac{1}{2\pi i}}\int_{c/p-i\infty}^{c/p+i\infty} t^{\varsigma}
	\bar\pzcQ_p(\varsigma) d\varsigma.
\end{equation}
	This is how far you can get using only Euler's series for the Riemann $\zeta$ function.

	Next we use that r.h.s.\Ref{eq:MellinRd}, understood as 
analytic extension of l.h.s.\Ref{eq:MellinRa}, is manifestly meromorphic in $\Cset$, having simple 
poles at $s=1$ (coming from the Riemann $\zeta$ function) and at ${s}/{p}= 2n-1,\,n\in\Nset$ (coming 
from the Euler $\Gamma$ function); the poles of the $\Gamma$ function at ${s}/{p}= 2(n-1),\,n\in\Nset$,
are ironed out by the pertinent zeros of the sine function.
	Note, though, that the $\zeta$ pole and the $\Gamma$ poles are located on different sides of the
fundamental strip.

	Therefore, if we now shift the contour in the $s$ plane to the right, beyond all $\Gamma$ poles, 
we obtain the Taylor series expansion of $\pzcS_p(t)$ about $t=0$, viz.
\begin{equation}
	\pzcS_p(t) 
=\label{eq:SexpandTAYLOR}
	\textstyle\sum\limits_{k=0}^{\infty}{\scriptstyle{(-1)^k}}\frac{\zeta(p[2k+1])}{(2k+1)!}t^{2k+1};\qquad\Re{p}>1.
\end{equation}
	It is readily checked that the same expansion is obtained directly from \Ref{eq:Sdef} by 
replacing $\sin(n^{-p}t)$ by its Maclaurin expansion, then exchanging the Maclaurin summation
with the summation over $n\in\Nset$ given in \Ref{eq:Sdef}, and using the Euler series of the Riemann
$\zeta$ function for $s>1$.

	If, on the other hand, we shift the contour to the left just a little bit
beyond the pole at $s=1$, say to $c_p =1-\eps$, we pick up the pole's residue and obtain
\begin{equation}
	\pzcS_p(t)  
=\label{eq:SpASYMPexpand} 
	\alpha_p^{} t^{1/p} + 
	{\textstyle\frac{1}{2\pi i}}\int_{\frac{c_p}{p}-i\infty}^{\frac{c_p}{p}+i\infty} t^{\varsigma}\,
	\bar\pzcQ_p(\varsigma)d\varsigma,
\end{equation}
with $\alpha_p^{}$ given in \Ref{eq:alphaSint}.
	Incidentally, by corollary to Theorem 1, the integral at r.h.s.\Ref{eq:SpASYMPexpand} is 
bounded in magnitude by $\beta_p|t|^{1/(p+1)}$.

	So we see how the Mellin transform plus the residue theorem of complex analysis reproduces --- in one elegant sweep ---
all the results we could establish with more elementary means, save the bound on the integral at 
r.h.s.\Ref{eq:SpASYMPexpand}. 
	Alas, with \Ref{eq:SexpandTAYLOR} and \Ref{eq:SpASYMPexpand} we exhaust the information 
about $\pzcS_p(t)$ which one can extract from the poles of $\bar\pzcQ_p(\varsigma)$.
	Clearly I cannot end on such a note!

	Let's see what we can learn from the fact that the fluctuations $\triangle\pzcS_p(t)$ 
about the trend $\alpha_p^{}t^{1/p}$ are given by the contour integral at r.h.s.\Ref{eq:SpASYMPexpand}.
	Proceeding now first formally, we pretend that we can shift the contour in \Ref{eq:SpASYMPexpand} 
to \emph{any} $c_p<1$.
	Writing it as $\{\varsigma=c_p/p +iv',v'\in\Rset\}$ and then
changing the integration variable in the contour integral at r.h.s.\Ref{eq:SpASYMPexpand} to $v'$, 
and then to $v=v'+ic_p/p$, gives 
\begin{eqnarray}
	\triangle\pzcS_p(t)
&=&\label{eq:DeltaSpINTrep} 
	{\textstyle\frac{1}{2\pi i}}\int_{c_p/p-i\infty}^{c_p/p+i\infty} t^{\varsigma}\,
	\bar\pzcQ_p(\varsigma)d\varsigma\\
&=& \label{eq:SpCONTOURintFOURIER} 
	{\textstyle{t^{c_p/p}}} \big(\cF^{-1} \bar{\pzcQ}_p({\textstyle\frac{c_p}{p}}+i\,\cdot\,)\big)(\ln t)\\
&=& \label{eq:SpCONTOURintFOURIERshifted} 
	\big(\cF^{-1} \bar{\pzcQ}_p(i\,\cdot\,)\big)(\ln t),
\end{eqnarray}
where 

$\phantom{nix}$
\vskip-1.2truecm
\begin{eqnarray}
	\big(\cF^{-1} f(\,\cdot\,)\big)(u) 
=\label{eq:FOURIERtransf} 
	{\textstyle\frac{1}{2\pi}}\int_{-\infty}^{\infty} e^{iv u} f(v)dv
\end{eqnarray}
denotes the formal inverse (non-unitary) Fourier transform of $f(v)$.
	To make proper sense out of the formal integral \Ref{eq:FOURIERtransf}, 
Fourier analysis enters next.

	Since the appropriate Fourier variable for $\triangle\pzcS_p(t),\, t>0$, is $u=\ln t$, Figs.11 and 12
are a bit deceptive now. 
	The graph of $u\mapsto \triangle\pzcS_p(e^u),\, u\in\Rset$, shown below for $p=2$, is a better guide to
one's intuition.
	The $u$ interval corresponds roughly to the $t$ interval in Fig.11, though not quite 
(note that $e^6\approx 400$).
	Fig.14 reveals that $\triangle\pzcS_2(e^u)$ has only one zero to the left of $u=0$ and 
vanishes $\asymp -\alpha_2 e^{u/2}$ when $u\downarrow -\infty$.
	On the other side of $u=0$, $\triangle\pzcS_2(e^u)$ develops oscillations with 
wavelengths which become exponentially small as $u$ gets large while their amplitudes grow with $u$, 
though bounded by the theoretical bounds $\pm\beta_2e^{u/3}$, with empirically optimized $\beta_2=20/29$ (cf. Fig.11).

\epsfxsize=9cm
\epsfysize=6cm
\centerline{\epsffile{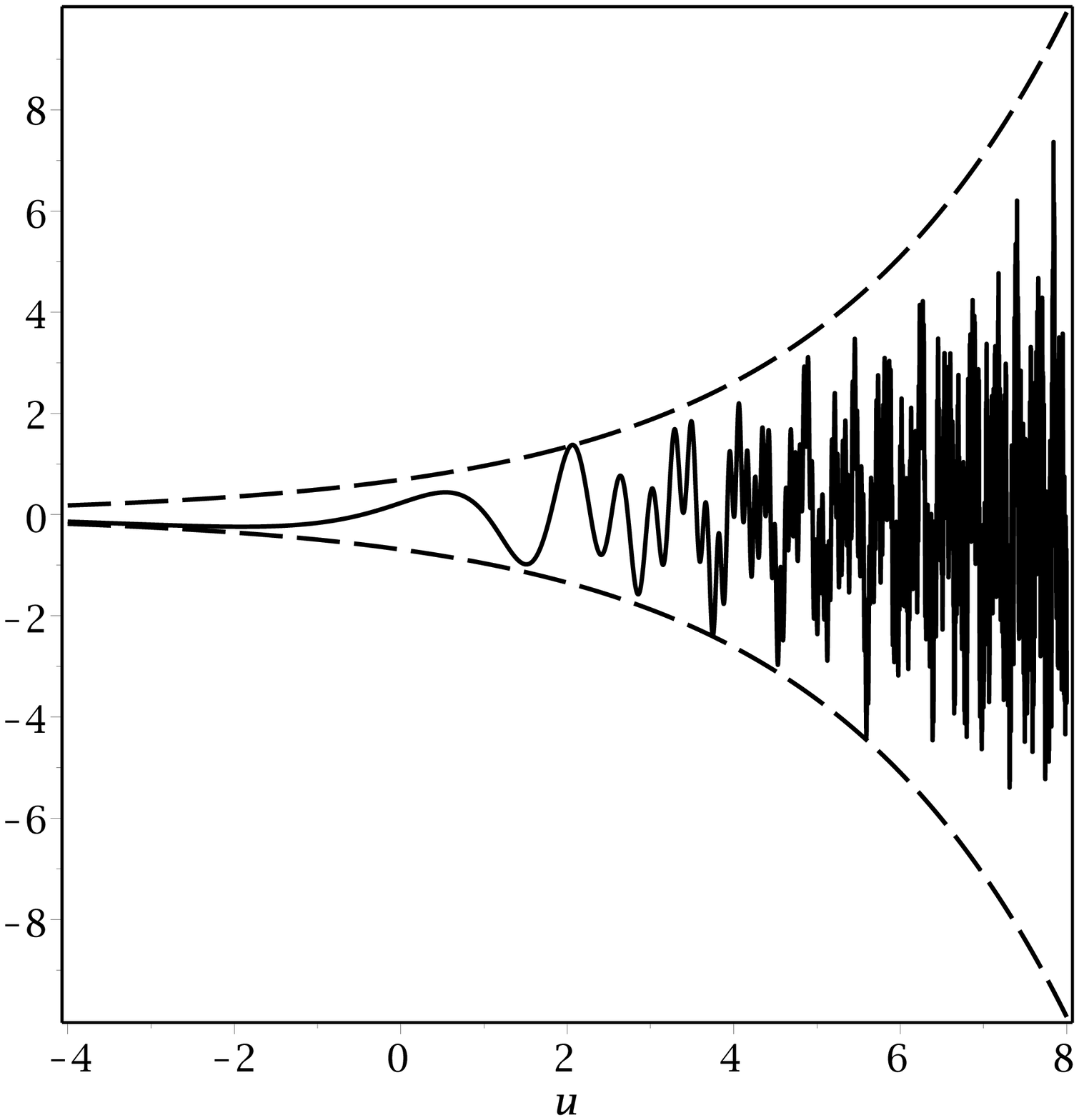}}
\begin{center}
\vskip-.6truecm
{$\qquad$\scriptsize{Fig.14. The graph of $u\mapsto \triangle\pzcS_2(e^u),\, u\in(-4,8)$,
		together with $\pm \frac{20}{29}\exp(u/3)$.}}
\end{center}
\vskip-.1truecm

\noindent

	What Fig.14 only hints at is that beyond $u=5.5$ the growth of the amplitudes departs more and more from our
theoretical bound.
	This is illustrated in Fig.15, which is the continuation of Fig.14 to the right ---
rescaled, of course, to fit on this page.

\epsfxsize=9.5cm
\epsfysize=6cm
\centerline{\epsffile{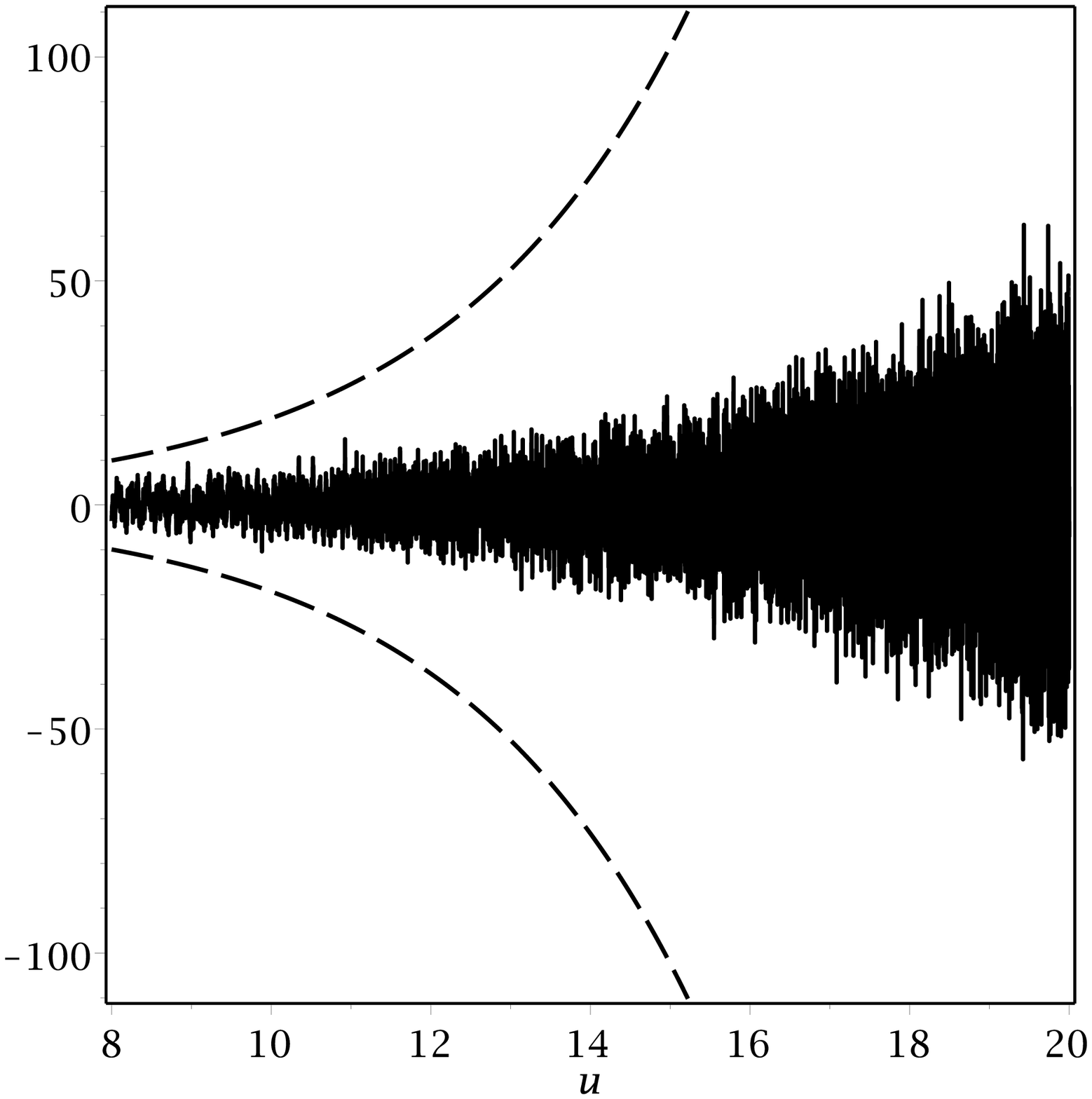}}
\begin{center}
\vskip-.5truecm
{$\qquad$\scriptsize{Fig.15. The graph of $u\mapsto \triangle\pzcS_2(e^u),\, u\in(8,20)$,
		together with $\pm \frac{20}{29}\exp(u/3)$.}}
\end{center}

\noindent
	Fig.15 leaves no doubt that the bounds $\pm\beta_2\exp(u/3)$ become lousy for large $u$;
in fact, something like $\pm\gamma_2\exp(u/5)$ traces the fluctuation amplitudes much better for $u\in(8,20)$,
but for even larger $u$ (not shown) also this bound will outgrow the fluctuations.
	(NB: According to the referees, see 4.1.1, the correct bound should be $\pm\kappa_2\exp(u/(6 +o(1)))$.)
	Nevertheless, as Fig.15 indicates, the fluctuations continue to grow forever.
	To prove their unbounded growth is not so easy, but
at least it is readily shown, for all $p>1$, that $\triangle\pzcS_p(e^u)$ does not approach $0$ when $u\to\infty$ ---
for suppose it would, then also $\triangle\pzcS_p(t)\to 0$ when $t\uparrow\infty$, and so then does its 
$t$ derivative (because $t^{(1-p)/p}\downarrow 0$ when $t\uparrow\infty$, and $\pzcS_p(t)$ contains a smallest wavelength); 
but the $t$ derivative of $\pzcS_p(t)$ is a manifestly quasi-periodic function of $t$: a contradiction --- end of proof.

	The upshot of this discussion is that $u\mapsto \triangle\pzcS_p(e^u)$ is a tempered distribution.
	Therefore, its  \emph{Fourier transform} 
$v\mapsto (\cF[\triangle\pzcS_p\circ\exp])(v) =\bar{\pzcQ}_p(iv)$
is to be understood \emph{in the sense of tempered distributions} as well.

	In this vein, let $\Ssp$ denote the Schwartz space of complex $\Csp^\infty$ functions on $\Rset$ 
which together with all their derivatives decay to zero at infinity faster than any power. 
	If $\psi\in\Ssp$, then its Fourier transform $\cF\psi\in\Ssp$, too, where 
\begin{equation}
	(\cF\psi)(v)
= \label{eq:FOURIERtDEF}
	\int_\Rset e^{-iuv}\psi(u)du.
\end{equation}
	The Fourier transform of a tempered distribution $g\in\Ssp^\prime$ is then defined by
\begin{equation}
	\int_\Rset (\cF g)(v)(\cF^{-1}\psi)(v)dv
= \label{eq:FOURIERtDEFtempered}
	\int_\Rset g(u)\psi(u)du\ \forall\ \psi\in\Ssp,
\end{equation}
where I hope to be forgiven for using the merely formal integral notation rather than a
proper dual pairing notation, cf. \cite{ReedSimonII}.

	As to the real function $u\mapsto \triangle\pzcS_p(e^u)$, for our purposes
it suffices to inspect its properties when integrated against the members of the family of shifted, 
scaled \emph{Hermite functions} $\{\psi_n\big(\kappa(u-w)\big)\in\Ssp: w\in\Rset,\kappa\in\Rset_+\}_{n=0}^\infty$, 
with
\begin{equation}
	\psi_n(u) 
= \label{eq:HERMITEfctns}
	(2^n n! \sqrt{\pi})^{-1/2} {e}^{-u^2/2} H_n(u) ,
\end{equation}
where
\begin{equation}
	H_n(u) 
= \label{eq:HERMITEpolynomials}
	(-1)^n {e}^{u^2} \textstyle\frac{d^n}{du^n} {e}^{-u^2}
\end{equation}
is the $n$-th Hermite polynomial, with $H_0\equiv 1$.
	In particular, to determine the late $t$ asymptotics of $\triangle\pzcS_p(t)$, we 
now define quantities of the form 
\begin{equation}
	A_n^p(w;\kappa)
:= \label{eq:TESTfunctionals}
	\int_\Rset \triangle\pzcS_p\big(e^u\big)\psi_n\big(\kappa(u-w)\big)du
\end{equation}
and evaluate their asymptotics as $w\to\infty$ with the help of \Ref{eq:FOURIERtDEFtempered}.

	Recalling that the Fourier transform
$\big(\cF[\triangle\pzcS_p\circ\exp]\big)(v)= \bar{\pzcQ}_p(iv)$, we have
\begin{equation}
	A_n^p(w;\kappa)
= \label{eq:TESTfunctionalsREWRITE}
	\int_\Rset \bar{\pzcQ}_p(iv) \big(\cF^{-1}[\psi_n\circ(\kappa(\,\cdot\,-w)]\big)(v)dv.
\end{equation}
	The integrand can be recast into a more convenient format by noting that
\begin{equation}
	\big(\cF^{-1}[\psi_n\circ(\kappa(\,\cdot\,-w)]\big)(v)
= \label{eq:FOURIERinvHERMITEshiftSCALED}
	e^{ivw} \textstyle\frac{1}{\kappa}\big(\cF^{-1}\psi_n\big)\big(\textstyle\frac{v}{\kappa}\big)
\end{equation}
and by recalling that the Hermite functions are $\Lsp^2$ eigenfunctions for $\cF$, viz.\footnote{The 
		factor $1/\sqrt{2\pi}$ is a consequence of working with the non-unitary version of $\cF$.}
\begin{equation}
	\big(\cF^{-1}\psi_n\big)(v)
= \label{eq:FOURIERinvHERMITE}
	\textstyle\frac{i^n}{\sqrt{2\pi}} \psi_n(v).
\end{equation}
	Note that $v\mapsto\psi_n(v)$ is even for even $n$ and odd for odd $n$.
	Furthermore, consulting \cite{Edwards} (or \cite{Titchmarsh}, \cite{Ivic}), one sees that it follows directly 
from the explicit formula \Ref{eq:QpMELLINdef} that\footnote{The Fourier transform 
		$v\mapsto \bar{\pzcQ}_p(iv)$ of the real function 
		$u\mapsto \triangle\pzcS_p(e^u)$ cannot be purely real or purely imaginary,
		for $\triangle\pzcS_p(e^u)$ is neither even nor odd, see Fig.14.}
$v\mapsto \Re\big(\bar{\pzcQ}_p(iv)\big)$ is even and 
$v\mapsto \Im\big(\bar{\pzcQ}_p(iv)\big)$ is odd, 
shown for $p=2$ in Fig.16:

\epsfxsize=9cm
\epsfysize=6cm
\centerline{\epsffile{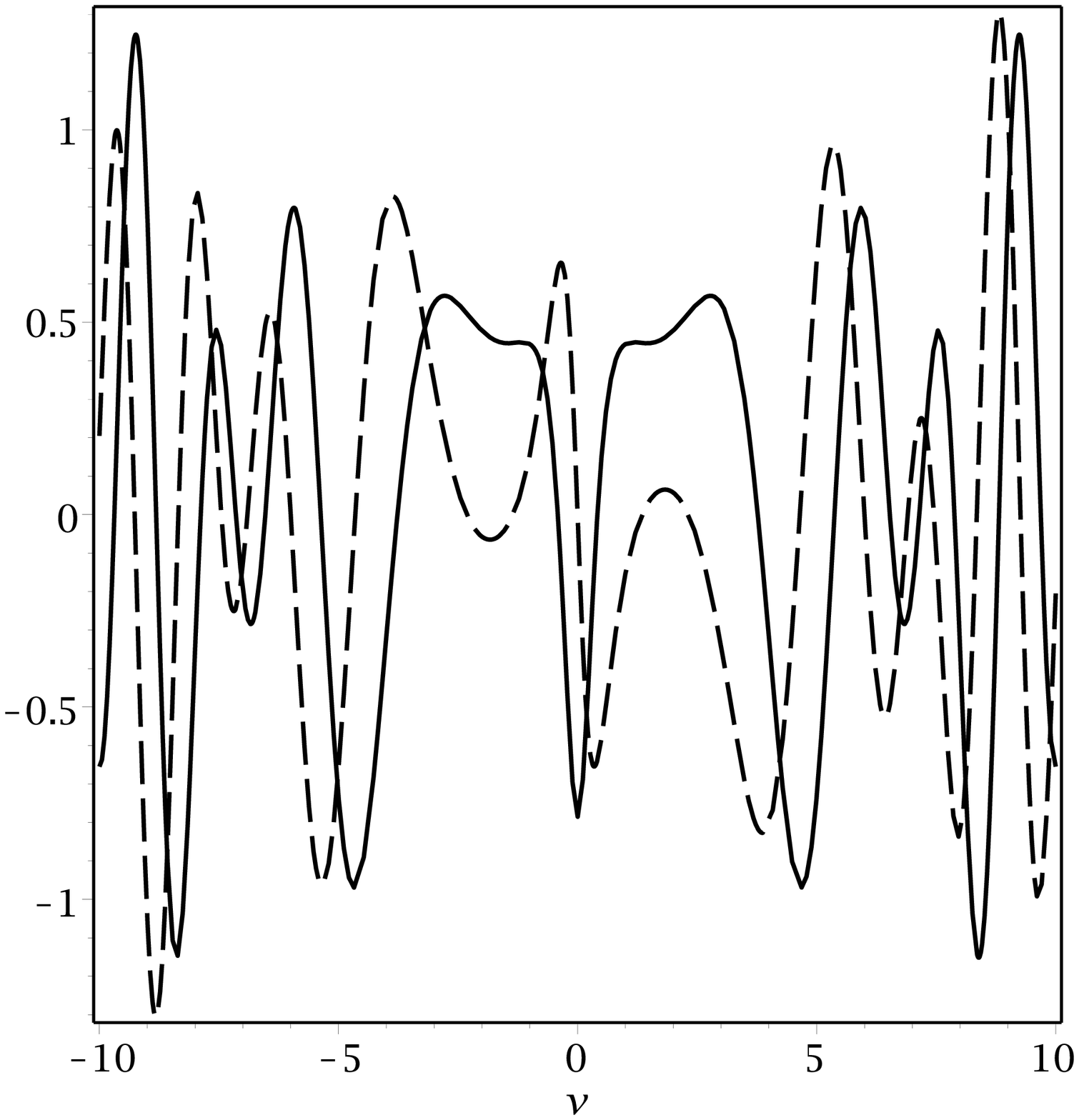}}
\begin{center}
\vskip-.6truecm
{$\qquad$\scriptsize{Fig.16. Real (solid) and imaginary (dashed) parts of $v\mapsto \bar{\pzcQ}_2(iv),\, v\in\Rset$.}}
\end{center}

\noindent
	Therefore, the real part of the function
$v\mapsto e^{ivw}\bar{\pzcQ}_p(iv)$ is even, while its imaginary part is odd. 
	In summary, we can conclude that \Ref{eq:TESTfunctionalsREWRITE} simplifies to 
\begin{eqnarray}
	A_n^p(w;\kappa)
\!\!&=&\!\! \label{eq:AnEVEN}
	(-1)^{\frac{n}{2}}	{\textstyle{\sqrt{\frac{2}{\pi}}}}
	\Re \int_{0}^{\infty} 
	\textstyle\frac{1}{\kappa}\psi_n\big(\textstyle\frac{v}{\kappa}\big) \bar\pzcQ_p(iv) e^{iv w}
		dv\quad\, (n\ \mathrm{even}),\\
&\mathrm{resp.}& \notag\\
	A_n^p(w;\kappa)
\!\!&=&\!\! \label{eq:AnODD}
	(-1)^{\frac{n+1}{2}}	{\textstyle{\sqrt{\frac{2}{\pi}}}}
	\Im \int_{0}^{\infty} 
	\textstyle\frac{1}{\kappa}\psi_n\big(\textstyle\frac{v}{\kappa}\big) \bar\pzcQ_p(iv) e^{iv w} 
	dv\quad (n\ \mathrm{odd}).
\end{eqnarray}
	The pertinent real or imaginary part of the integrand which features in the integrals at 
r.h.s.\Ref{eq:AnEVEN},\Ref{eq:AnODD} is shown  for $n=0$ and $n=1$ in Figs.17 \& 18 below, respectively,
in each case with  $p=2$, $\kappa=1$, and $w=60$. 
\smallskip

\epsfxsize=9cm
\epsfysize=6cm
\centerline{\epsffile{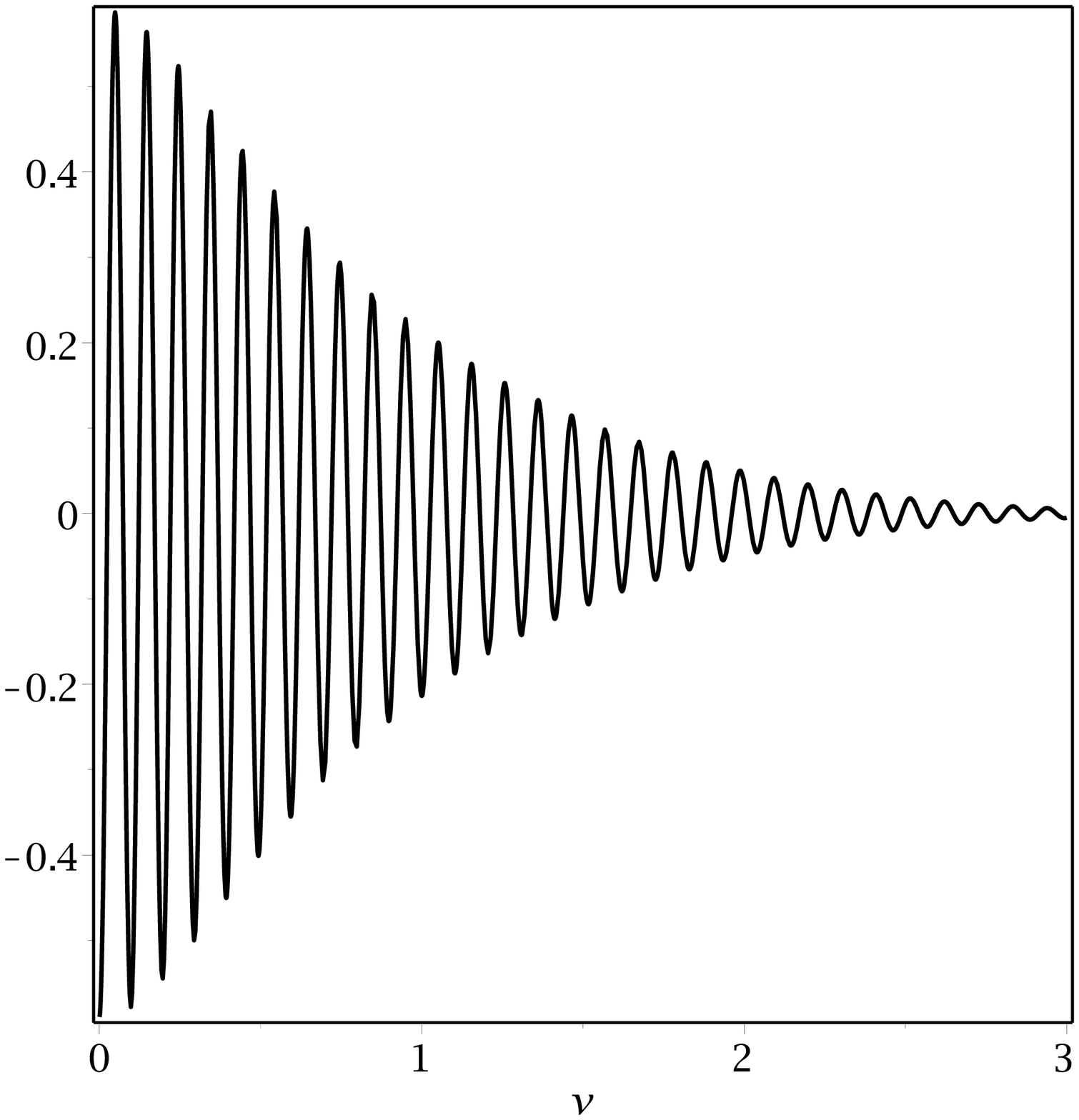}}
\begin{center}
\vskip-.5truecm
{$\qquad$\scriptsize{Fig.17. The graph of $\Re\big(e^{iv w} \bar\pzcQ_2(iv)\big) \psi_0(v)$ for $w=60$.}}
\end{center}
\newpage

\epsfxsize=9cm
\epsfysize=6cm
\centerline{\epsffile{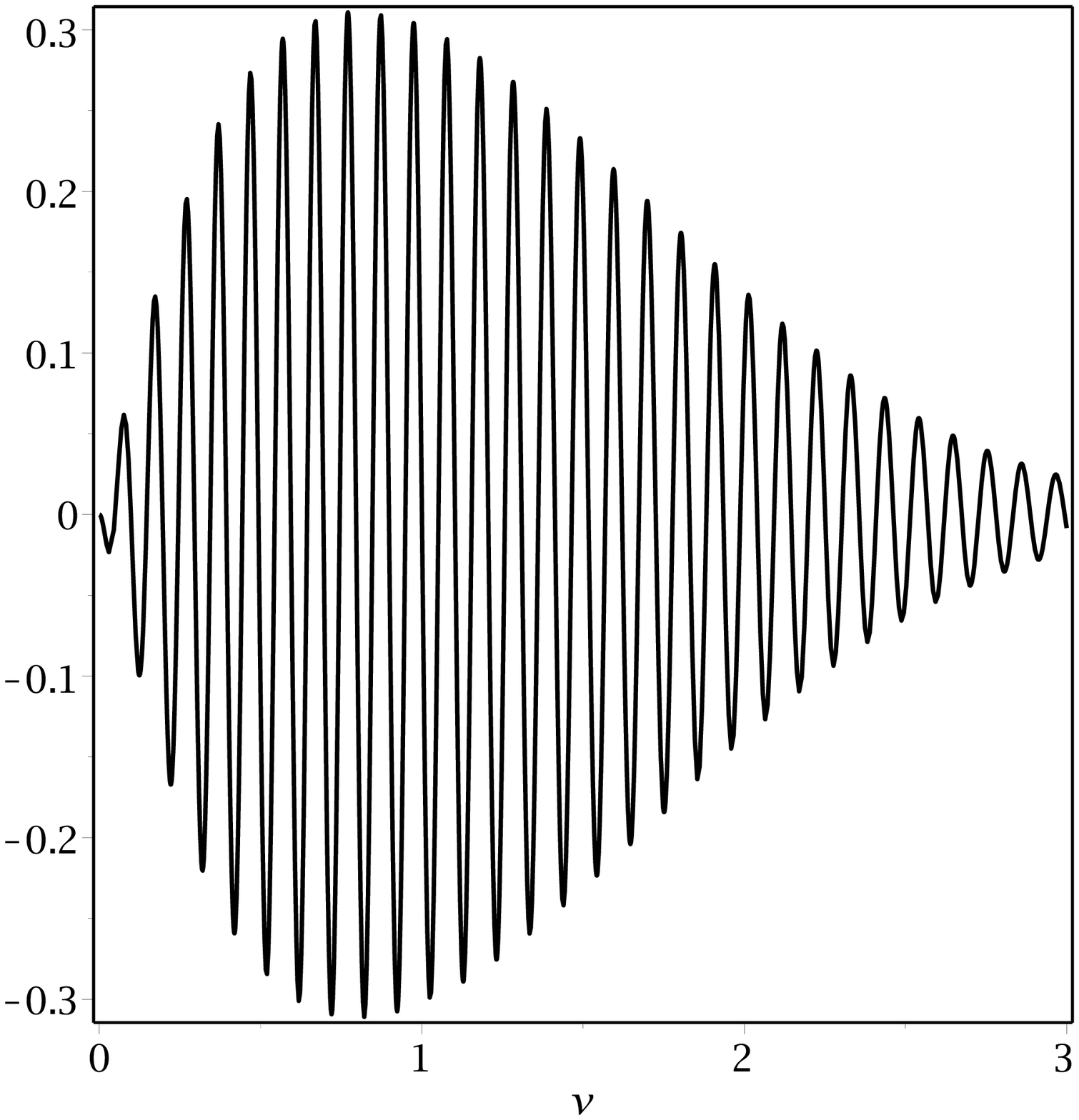}}
\begin{center}
\vskip-.5truecm
{$\qquad$\scriptsize{Fig.18. The graph of $\Im\big(e^{iv w} \bar\pzcQ_2(iv)\big) \psi_1(v)$ for $w=60$.}}
\end{center}

\noindent
	Since for each choice of $n$, $p$, $\kappa$ the integrand of the integrals at 
r.h.s.\Ref{eq:AnEVEN},\Ref{eq:AnODD} becomes a highly oscillatory Schwartz function, by the 
Riemann--Lebesgue lemma the $A^p_n(w;\kappa)$ vanish in the limit $w\to\infty$, and 
we are interested in how they vanish asymptotically when $w\to\infty$. 
	Since the integrand is a complex analytic function in a strip neighborhood of $\Rset$, 
their $w\to\infty$ asymptotic behavior is easily found from the full asymptotic expansion
\begin{eqnarray}
\hskip-.5truecm
	\int_{0}^{\infty}\!\!
	\textstyle\frac{1}{\kappa}\psi_n\big(\textstyle\frac{v}{\kappa}\big) \bar\pzcQ_p(iv) e^{iv w} 
	dv
\!\!\!&=&\!\!\!\! \label{eq:AnpINTasympEXPANSION}
	\textstyle{ 
	\sum\limits_{\ell\in\Nset}^{} \left[\frac{2\pi}{w}\right]^\ell 
	\!\frac{d^\ell}{dv^\ell}\! \left[
	\textstyle\frac{1}{\kappa}\psi_n\big(\textstyle\frac{v}{\kappa}\big) \bar\pzcQ_p(iv)
				\right]\!(0)
	\sum\limits_{j=0}^\ell \frac{1}{(\ell-j)!}\!\left[\frac{i}{2\pi}\right]^{j+1}}\!\!\!.
\end{eqnarray}
	(Here, $[\cdots]$ does not mean ``integer part.'')
	Inserting \Ref{eq:AnpINTasympEXPANSION} at r.h.s.\Ref{eq:AnEVEN},\Ref{eq:AnODD} yields the aysmptotic $w\to\infty$ 
expansion of $A_n^p(w;\kappa)$. 
	For most practical matters we will only need the pertinent lowest order nonvanishing
term in \Ref{eq:AnpINTasympEXPANSION}, as in the example below.
\vskip-.5truecm

%%%%%%%%%%%%%%%%%%%%%%%%%%%%%%%%%%%%%%%%%%%%%%%%%%%%%%%%%%%%%%%%%%%%%%%%%%%%%%%%%%%%%%%%%
%%%%%%%%%%%%%%%%%%%%%%%%%%%%%%%%%%%%%%%%%%%%%%%%%%%%%%%%%%%%%%%%%%%%%%%%%%%%%%%%%%%%%%%%%
%%%%%%%%%%%%%%%%%%%%%%%%%%%%%%%%%%%%%%%%%%%%%%%%%%%%%%%%%%%%%%%%%%%%%%%%%%%%%%%%%%%%%%%%%
\subsubsection{Application: Gaussian averages of $u\mapsto \triangle\pzcS_p(e^u)$}
%%%%%%%%%%%%%%%%%%%%%%%%%%%%%%%%%%%%%%%%%%%%%%%%%%%%%%%%%%%%%%%%%%%%%%%%%%%%%%%%%%%%%%%%%
%%%%%%%%%%%%%%%%%%%%%%%%%%%%%%%%%%%%%%%%%%%%%%%%%%%%%%%%%%%%%%%%%%%%%%%%%%%%%%%%%%%%%%%%%
%%%%%%%%%%%%%%%%%%%%%%%%%%%%%%%%%%%%%%%%%%%%%%%%%%%%%%%%%%%%%%%%%%%%%%%%%%%%%%%%%%%%%%%%%
	I briefly register the bonus of the asymptotics obtained in the previous subsection: 
since $\psi_0(u) = \pi^{-1/4}e^{-u^2/2}$, it follows that 
$A_0^p(w;1)$ is proportional to a standard Gaussian average of $\triangle\pzcS_p(e^u)$ centered at $w$.
	For instance, 
\begin{eqnarray}
	A_0^2(w;1) 
= \label{eq:AnullZWEIeinsOFw}
	\textstyle{\frac{\pi^{1/4}}{\surd{2}} \big(\frac12 \gamma+\ln(2\pi) \big) \frac{1}{w} + 
	O\left(\frac{1}{w^2}\right)},
\end{eqnarray}
where $\gamma = 0.57721...$ is Euler's constant.
	Multiplication of \Ref{eq:AnullZWEIeinsOFw} by $\pi^{1/4}/\sqrt{2\pi}$ yields the standard Gaussian 
average of $\triangle\pzcS_2(e^u)$ centered at $w$.

	The road is now paved to analyze the fluctuations of $u\mapsto \triangle\pzcS_p(e^u)$ as $w\to\infty$.
	If one thinks of the $w$-centered Gaussian $u$-average as an analogue of Kac's uniform $t$-average, 
one can ask how much ``$u$-time'' the function $\triangle\pzcS_p(e^u)$ spends on average, centered at $w$, in
some value interval $(a,b)$.
	Markov's method can be applied, and one might even be able to compute the answer 
explicitly using the formalism of the previous subsection, whether the answer is ``normal'' or not. 
	So much for the Mellin transform.
%\newpage

%%%%%%%%%%%%%%%%%%%%%%%%%%%%%%%%%%%%%%%%%%%%%%%%%%%%%%%%%%%%%%%%%%%%%%%%%%%%%%%%%%%%%%%%%
%%%%%%%%%%%%%%%%%%%%%%%%%%%%%%%%%%%%%%%%%%%%%%%%%%%%%%%%%%%%%%%%%%%%%%%%%%%%%%%%%%%%%%%%%
%%%%%%%%%%%%%%%%%%%%%%%%%%%%%%%%%%%%%%%%%%%%%%%%%%%%%%%%%%%%%%%%%%%%%%%%%%%%%%%%%%%%%%%%%
%%%%%%%%%%%%%%%%%%%%%%%%%%%%%%%%%%%%%%%%%%%%%%%%%%%%%%%%%%%%%%%%%%%%%%%%%%%%%%%%%%%%%%%%%
\section{Epilogue}
%%%%%%%%%%%%%%%%%%%%%%%%%%%%%%%%%%%%%%%%%%%%%%%%%%%%%%%%%%%%%%%%%%%%%%%%%%%%%%%%%%%%%%%%%
%%%%%%%%%%%%%%%%%%%%%%%%%%%%%%%%%%%%%%%%%%%%%%%%%%%%%%%%%%%%%%%%%%%%%%%%%%%%%%%%%%%%%%%%%
%%%%%%%%%%%%%%%%%%%%%%%%%%%%%%%%%%%%%%%%%%%%%%%%%%%%%%%%%%%%%%%%%%%%%%%%%%%%%%%%%%%%%%%%%
%%%%%%%%%%%%%%%%%%%%%%%%%%%%%%%%%%%%%%%%%%%%%%%%%%%%%%%%%%%%%%%%%%%%%%%%%%%%%%%%%%%%%%%%%

	After my presentation, which ended with sect. 4.1, a young participant at the meeting came to talk to me.
	He was incredulous, and something close to the following conversation ensued:\footnote{Many of the historical 
	developments and anecdotes I wrote up for this talk have been faithfully reconstructed from email
	records, but some of it only from my memory, and so it is appropriate to quote another of Jerry's favorites: 
	``Don't mistake me for the facts!''}

``Prof. Kiessling:

	Why are you doing this? What does this have to do with physics?''
\medskip

``Well, I am doing this because it's fun! And didn't I myself at the end of 

	the talk raise the question whether there are connections to physics? Also, 

	recall that I noted that the %Bohr 
	Schr\"odinger hydrogen spectrum has eigenvalues 

	$-n^{-2}$, so $n^{-2}$ frequencies do occur in physics, you get $\cos(n^{-2}t)+i\sin(n^{-2}t)$ 

	as $t$-dependent factors in any expansion using the eigenwave functions.
	Of 

	course, the series $\pzcS_2(t)$ itself may not occur.''
\medskip

``Exactly, how can you hope that somehow this will be useful to physics?

	This is crazy!''
\medskip

	Truth be told, I am not sure he really said ``crazy,'' but he certainly gave me the impression that the thought
crossed his mind. 
	And on this anecdote I close by quoting Jerry Percus from an interview he gave a few years ago at the
Courant Institute \cite{Ball}:
\medskip

	``What you want is to be a little bit crazy. You want to think of things 

	that sound like nonsense to start with and then when you get deeper, 

	they're not nonsense at all.''
\smallskip

\centerline{***}

\vfill\vfill

\noindent
\textbf{Acknowledgement:} 
I thank Steve Greenfield for his original question which triggered this line of inquiry;
I also wish him a happy retirement. 
Next I thank Jared Speck and Mikko Stenlund for their participation in nailing down $\alpha_2^{}$.
Thanks go to Norm Frankel and Steve Miller for their enlightening explanations of the relationship 
between $\pzcS_p(t)$ and $\zeta(s)$ after my SMM~106 talk;  for further enlightenment see also
\cite{FrankelETal,FrankelETalB}.
Of course, I thank all of these, and also Michael Fisher, for allowing me to quote from their emails
they had sent me.
I also thank Joel Lebowitz for the honorable invitation to contribute to SMM 106 and to this special issue, 
and the three anonymous referees for their generously offered expertise in analytical number theory (including
11 references, amongst them  also \cite{HardyLittlewood,Flett}) which allowed me to revise the perspective on 
$\pzcS_p(t)$ presented in this paper; it also allowed me to correct my Conjecture 1. 
But most of all, I thank all three honorees for all they taught (all of) us at so many a statistical mechanics meeting, 
and Jerry Percus in particular --- for being a wonderful mentor and friend!

\newpage

%%%%%%%%%%%%%%%%%%%%%%%%%%%%%%%%%%%%%%%%%%%%%%%%%%%%%%%%%%%%%%%%%%%%%%%%%%%%%%%%%%%%%%%%%
%%%%%%%%%%%%%%%%%%%%%%%%%%%%%%%%%%%%%%%%%%%%%%%%%%%%%%%%%%%%%%%%%%%%%%%%%%%%%%%%%%%%%%%%%
%%%%%%%%%%%%%%%%%%%%%%%%%%%%%%%%%%%%%%%%%%%%%%%%%%%%%%%%%%%%%%%%%%%%%%%%%%%%%%%%%%%%%%%%%
%%%%%%%%%%%%%%%%%%%%%%%%%%%%%%%%%%%%%%%%%%%%%%%%%%%%%%%%%%%%%%%%%%%%%%%%%%%%%%%%%%%%%%%%%
\section*{Appendix}
%%%%%%%%%%%%%%%%%%%%%%%%%%%%%%%%%%%%%%%%%%%%%%%%%%%%%%%%%%%%%%%%%%%%%%%%%%%%%%%%%%%%%%%%%
%%%%%%%%%%%%%%%%%%%%%%%%%%%%%%%%%%%%%%%%%%%%%%%%%%%%%%%%%%%%%%%%%%%%%%%%%%%%%%%%%%%%%%%%%
%%%%%%%%%%%%%%%%%%%%%%%%%%%%%%%%%%%%%%%%%%%%%%%%%%%%%%%%%%%%%%%%%%%%%%%%%%%%%%%%%%%%%%%%%
%%%%%%%%%%%%%%%%%%%%%%%%%%%%%%%%%%%%%%%%%%%%%%%%%%%%%%%%%%%%%%%%%%%%%%%%%%%%%%%%%%%%%%%%%

\vskip-.3truecm
	Here we fulfill Greenfield's request and produce lower and upper bounds on $\pzcS_2(t)$ suitable
for an undergraduate workshop.

	First of all, pick $N\in\Nset\cup\{0\}$ and split r.h.s.\Ref{eq:Sdef} with $p=2$ into two parts,
\begin{equation}
	\pzcS_{2}(t) 
=\label{eq:SdefZWEIsplit}
	\textstyle\sum\limits_{n=1}^{N}\sin(n^{-{2}}t)
	+
	\textstyle\sum\limits_{n=N+1}^{\infty}\sin(n^{-{2}}t),
\end{equation}
with $\sum_{n=1}^0\equiv 0$.
	Now $\sin(\xi)\geq -1$ estimates the first sum from below by $-N$, and
$\sin(\xi)\geq \xi - \xi^3/6$ for $\xi\geq 0$ is used to bound the second sum by 
\begin{equation}
	\textstyle\sum\limits_{n=N+1}^{\infty}\sin(n^{-{2}}t)
\geq \label{eq:SdefZWEIsplitPARTtwoLOWERboundN}
	\Big(\textstyle\sum\limits_{n=N+1}^{\infty} n^{-{2}}\Big)t
	-
	\textstyle\frac{1}{6}\Big(\sum\limits_{n=N+1}^{\infty}n^{-{6}}\Big)t^3.
\end{equation}
	The converging $p$-series are easy to evaluate when $N$ is not too big, using
\begin{equation}
	\textstyle\sum\limits_{n=N+1}^{\infty} n^{-{2}}
= \label{eq:AN}
	\frac{\pi^2}{6} - \textstyle\sum\limits_{n=1}^N n^{-{2}}\,,
\end{equation}
\begin{equation}
	\textstyle\sum\limits_{n=N+1}^{\infty}n^{-{6}}
= \label{eq:BN}
	\frac{\pi^6}{945} - \sum\limits_{n=1}^{N}n^{-{6}}\,.
\end{equation}
	And so, with the abbreviations
r.h.s.\Ref{eq:AN}$\equiv A_2(N)$ and $\frac{1}{6}\times$ r.h.s.\Ref{eq:BN}$\equiv B_2(N)$, 
\begin{equation}
	\pzcS_{2}(t) 
\geq \label{eq:SdefZWEIsplitLOWERboundN}
	- N
	+
	A_2(N) t
	-
	B_2(N) t^3\qquad \forall\ N\in\Nset\cup\{0\},
\end{equation}
and thus
\begin{equation}
	\pzcS_{2}(t) 
\geq \label{eq:SdefZWEIsplitLOWERboundMAXoverN}
	\max\{- N + A_2(N) t - B_2(N) t^3:\, 0\leq N\leq N_* \}\quad \forall\ N_*\in\Nset\cup\{0\}.
\end{equation}
	It suffices to pick $N_*=7$ to obtain a ready-to-plot small family of cubic parabolas, 
the pointwise maximum of which is a positive lower bound to $\pzcS_2(t)$ which ``tilts upward'' 
over the whole interval $0<t<120$.

	For the sake of completeness of the discussion I should
note that  with equal ease one can also produce a complementary upper bound to $\pzcS_2(t)$.
	Namely, instead of $\sin(\xi)\geq -1$ one now uses $\sin(\xi)\leq 1$ to estimate the
first sum in \Ref{eq:SdefZWEIsplit} from above by $N$, next one uses that 
$\sin(\xi)\leq \xi - \xi^3/6 + \xi^5/120$ for $\xi\geq 0$ to estimate the second sum in \Ref{eq:SdefZWEIsplit} from 
above by 
\begin{equation}
	\textstyle\sum\limits_{n=N+1}^{\infty}\sin(n^{-{2}}t)
\leq \label{eq:SdefZWEIsplitPARTtwoUPPERboundN}
	\Big(\textstyle\sum\limits_{n=N+1}^{\infty} n^{-{2}}\Big)t
	-
	\textstyle\frac{1}{6}\Big(\sum\limits_{n=N+1}^{\infty}n^{-{6}}\Big)t^3
	+
	\textstyle\frac{1}{120}\Big(\sum\limits_{n=N+1}^{\infty}n^{-{10}}\Big)t^5.
\end{equation}
	The first two converging $p$-series are just the same as before, the third one is new,
but which is equally easy to evaluate when $N$ is not too big, using
\begin{equation}
	\textstyle\sum\limits_{n=N+1}^{\infty} n^{-{10}}
= \label{eq:CN}
	\frac{\pi^{10}}{93555} - \textstyle\sum\limits_{n=1}^N n^{-{10}}.
\end{equation}
	And so, with the new abbreviation $\frac{1}{120}\times$ r.h.s.\Ref{eq:CN}$\equiv C_2(N)$, 
we have 
\begin{equation}
	\pzcS_{2}(t) 
\leq \label{eq:SdefZWEIsplitUPPERboundN}
	- N
	+
	A_2(N) t
	-
	B_2(N) t^3
	+
	C_2(N) t^5 \qquad \forall\ N\in\Nset\cup\{0\},
\end{equation}
and thus, $\forall\ N_*\in\Nset\cup\{0\}$, 
\begin{equation}
	\pzcS_{2}(t) 
\leq \label{eq:SdefZWEIsplitUPPERboundMINoverN}
	\min\{- N + A_2(N) t - B_2(N) t^3 + C_2(N) t^5:\, 0\leq N\leq N_* \}. 
\end{equation}
	Once again picking $N_*=7$ we now obtain a ready-to-plot small family of quintic polynomials, 
the pointwise minimum of which is a positive upper bound to $\pzcS_2(t)$ which also ``tilts upward'' 
over the whole interval $0<t<120$.
	Together with the family of lower bounds from above, this does produce an upward tilted corridor
in which the graph of $\pzcS_2(t)$ must lie.
	This is illustrated in Fig.19.

\epsfxsize=10.5cm
\epsfysize=7cm
\centerline{\epsffile{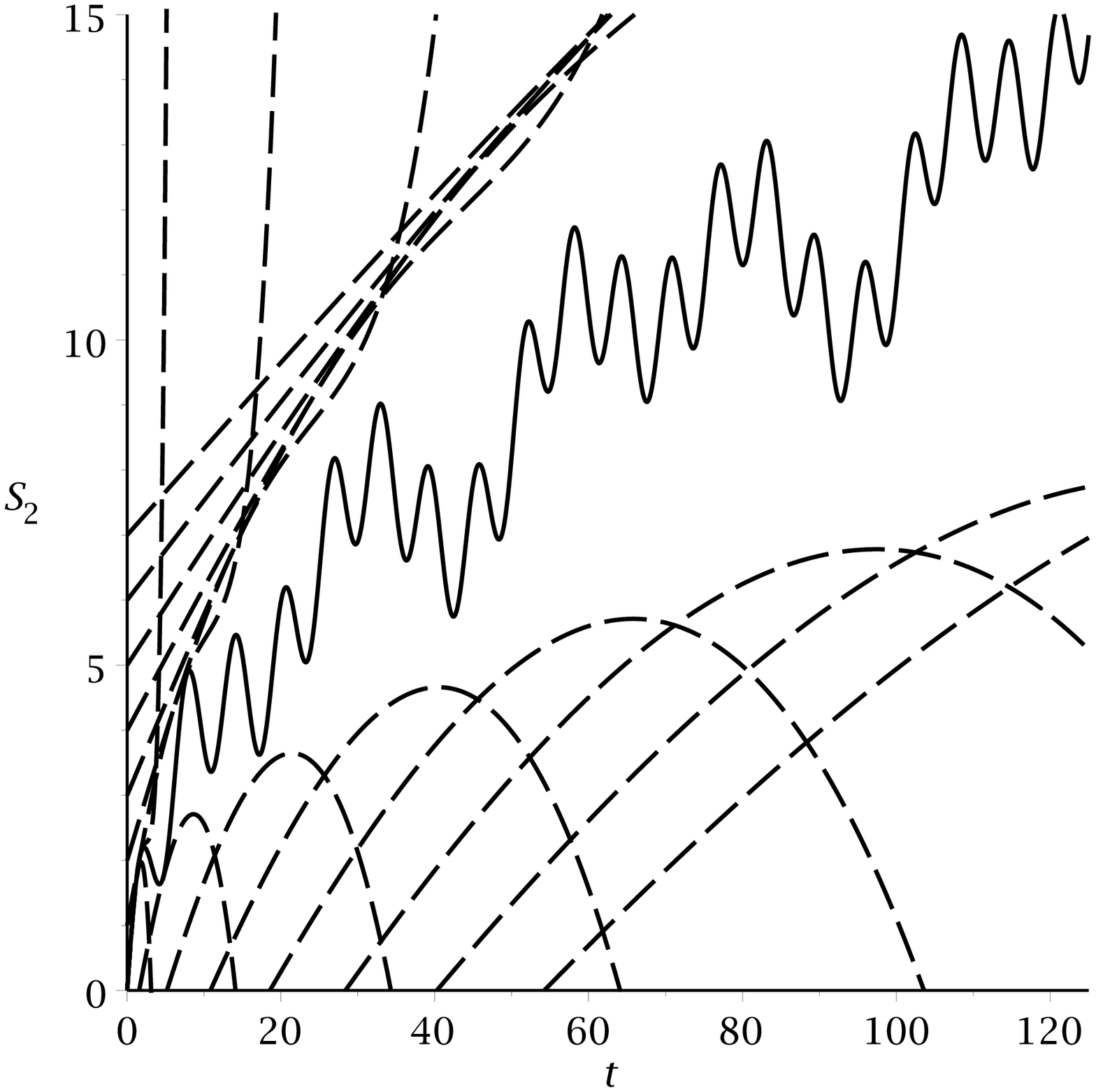}}
\begin{center}
\vskip-.3truecm
{$\qquad$\scriptsize{Fig.19. The 2,000-th partial sum of $\pzcS_2(t)$ 
	together with the bounds \Ref{eq:SdefZWEIsplitLOWERboundN} and
	\Ref{eq:SdefZWEIsplitUPPERboundN} for $0\leq N\leq 7$.}}
\end{center}

%%%%%%%%%%%%%%%%%%%%%%%%%%%%%%%%%%%%%%%%%%%%%%%%%%%%%%%%%%%%
%%%%%%%%%%%%%%%%%%%%%%%%%%%%%%%%%%%%%%%%%%%%%%%%%%%%%%%%%%%%
%%%%%%%%%%%%%%%%%%%%%%%%%%%%%%%%%%%%%%%%%%%%%%%%%%%%%%%%%%%%
%%%%%%%%%%%%%%%%%%%%%%%%%%%%%%%%%%%%%%%%%%%%%%%%%%%%%%%%%%%%

%

\small
\begin{thebibliography}{9999999}\setlength{\parskip}{5pt}
\bibitem[Bal08]{Ball}
Ball, M.L.,
	``After 50 years at Courant, Jerry Percus is still `A little crazy',''
	p.3 in: \textit{Courant Inst. of Math. Sci. Newsletter}, Fall/Winter 2008.
\vskip-.35truecm
\noindent
\bibitem[Ble92]{Bleher}
Bleher, P.,
	``On the distribution of the number of lattice points inside a family of convex ovals,''
	\textit{Duke Math. J.} \textbf{67}:461--481 (1992).
\vskip-.35truecm
\noindent
\bibitem[BCDL93]{BleherETal}
Bleher, P.,
Cheng, Z.,
Dyson, F.J.,
and
Lebowitz, J.L.,
	``Distribution of the error term for the number of lattice points inside a shifted circle,''
	\textit{Commun. Math. Phys.} \textbf{154}:433--469 (1993).
\vskip-.35truecm
\noindent
\bibitem[ChUb07]{ChUb}
Chamizo, F.,
and 
Ubis, A.,
	``Some Fourier series with gaps,''
	\textit{J. Anal. Math.} \textbf{101}:179--197 (2007).
\vskip-.35truecm
\noindent
\bibitem[Dys72]{Dyson}
Dyson, F.
	``Missed opportunities,''
	\textit{Bull. Amer. Math. Soc.} \textbf{78}:635--652 (1972).
\vskip-.35truecm
\noindent
\bibitem[Edw74]{Edwards}
Edwards, H.M.,
	\emph{Riemann's $\zeta$ function},
	Dover, Mineola, NY (1974).
\vskip-.35truecm
\noindent
\bibitem[Fle50]{Flett}
Flett, T.M.,
	``On the function $\sum_{n=1}^\infty (1/n)sin(t/n)$,''
	\textit{J. London Math. Soc.} \textbf{25}:5--19 (1950).
\vskip-.35truecm 
\noindent
\bibitem[GrKo91]{expSUMSbook}
Graham, S.W.,
and
Kolesnik, G.,
	\textit{Van der Corput's method of exponential sums},
	London Math. Soc. Lect. Note Series \textbf{126},
	Cambridge Univ. Press (1991).
\vskip-.35truecm
\noindent
\bibitem[HaLi36]{HardyLittlewood}
Hardy, G.H., 
and
Littlewood, J.E.,
	``Notes on the theory of series (XX): On Lambert series,''
	\textit{Proc. London Math. Soc.} (Ser.2) \textbf{41}:257--270 (1936).
\vskip-.35truecm 
\noindent
\bibitem[HeBr92]{HeathBrown}
Heath-Brown, D.R,
	``The distribution and moments of the error term in the Dirichlet divisor problem,'' 
	Acta Arithm. \textbf{60}:389--414 (1992).
\vskip-.35truecm
\noindent
\bibitem[Ivi03]{Ivic}
Ivi\'c, A.,
	\emph{The Riemann $\zeta$ function},
	Dover, Mineola, NY (2003).
\vskip-.35truecm
\noindent
\bibitem[IwKo04]{IwaKowa}
Iwaniec, H., 
and
Kowalski, E.,
	\textit{Analytic number theory},
	Amer. Math. Soc. Colloquium Pub. \textbf{53},
	Amer. Math. Soc., Providence, RI (2004).
\vskip-.35truecm 
\noindent
\bibitem[Kac38]{KacFRENCH}
Kac, M.,
	``{Sur les fonctions ind\'ependantes V},''
	\textit{Studia Math.} \textbf{7}:96--100 (1938). %NB: often referred to as 1937 (that's the submission year)
\vskip-.35truecm 
\noindent
\bibitem[Kac43]{KacAJM}
Kac, M.,
	``{On the distribution of values of trigonometric sums with linearly independent frequencies},''
	\textit{Amer. J. Math.} \textbf{65}:609--615 (1943).
\vskip-.35truecm 
\noindent
\bibitem[Kac59]{KacBOOK}
Kac, M.,
	\emph{Statistical independence in probability, analysis, and number theory},
	Math. Assoc. Amer. (Publ.), John Wiley \& Sons (Distr.) (1959).
\vskip-.35truecm 
\noindent
\bibitem[Kie08]{KiesslingASSISI}
Kiessling, M.K.-H.,
	``{Statistical equilibrium dynamics,}''
        pp.91-108 in \textit{AIP Conf. Proc.} \textbf{970},
        A. Campa, A. Giansanti, G. Morigi, and F. Sylos Labini (eds.), American Inst. Phys. (2008).
\vskip-.35truecm
\noindent
\bibitem[KFGT95]{FrankelETalB}
Kowalenko, V.,
Frankel, N.E., 
Glasser, M.L.,
and
Taucher, T.,
	``Generalised Euler-Jacobi inversion formula and asymptotics beyond all orders,''
	\textit{London Math. Soc. Lect. Note Ser.} \textbf{214}, Cambridge Univ. Press, Cambridge (1995).
\vskip-.35truecm
\noindent
\bibitem[McK90]{McKean}
McKean, H.P.,
	``Mark Kac: 1914--1984,''
	\emph{A biographical memoir}, 
	Publ. of the Nat. Acad. Sci., Washington (1990).
\vskip-.35truecm
\noindent
\bibitem[MiSch04]{MillerSchmid}
Miller, S.D.,
and 
Schmid, W.,
	``The highly oscillatory behavior of automorphic distributions for $SL(2)$,'' 
	\textit{Lett. Math. Phys.} \textbf{69}:265-286 (2004).
		% Preprint http://www.math.harvard.edu/schmid/.
\vskip-.35truecm
\noindent
\bibitem[Mon94]{Montgomery}
Montgomery, H.L.,
	\textit{Ten lectures on the interface between analytic number theory and harmonic analysis}, 
	CBMS Regional Conf. Series in Math. \textbf{84},
	% Published for the Conference Board of the Mathematical Sciences, 
	Washington, DC (1994).
\vskip-.35truecm
\noindent
\bibitem[NHFG92]{FrankelETal}
Ninham, B.W., 
Hughes, B.D.,
Frankel, N.E., 
and
Glasser, M.L.,
	``M\"obius, Mellin, and mathematical physics,''
	\textit{Physica A} \textbf{186}:441--481 (1992).
\vskip-.35truecm
\noindent
\bibitem[Tit86]{Titchmarsh}
Titchmarsh, E.C.,
	\textit{The theory of the Riemann zeta-function}, $2^{nd}$ed.,
	edited and with a preface by D. R. Heath-Brown.
	Clarendon Press, New York (1986). 
\vskip-.35truecm
\noindent
\bibitem[ReSi75]{ReedSimonII}
Reed, M.,
and
Simon, B.,
	\emph{Fourier Analysis, Self-Adjointness}, ``Methods of modern mathematical physics, II,''
	Acad. Press, San Diego (1975).
\vskip-.35truecm
\noindent
\bibitem[Rie76]{RiemannWERKE}
Riemann, G.F.B.,
	\emph{Gesammelte mathematische Werke und wissen\-schaft\-licher Nachlass},
	R. Dedekind, H. Weber (eds.),
	Teubner, Leipzig (1876).
\vskip-.35truecm
\noindent
\bibitem[Vin04]{Vinogradov}
Vinogradov, I.M.,
	\textit{The method of trigonometrical sums in the theory of numbers},
	Dover Publications Inc., Mineola, NY, 2004. 
	% Translated from the Russian, revised and annotated by K. F. Roth and Anne Davenport,
	% Reprint of the 1954 translation.
\vskip-.35truecm
\noindent
\bibitem[Zyg02]{Zygmund}
\hskip-2pt
Zygmund, A.,
	\emph{Trigonometric series}, $3^{rd}$ed, Cambridge Univ. Press~(2002).
\end{thebibliography}
\end{document}